\documentclass[twocolumn]{aastex63}
\shorttitle{Refractory-to-Volatile Ratios}
\shortauthors{Lothringer et al.}

\usepackage{amsmath}
\usepackage{xspace}
\usepackage{float}

\def\mearth{{\rm\,M_\oplus}}

\newcommand{\microns}{$\mu$m}

\newcommand{\mjup}{${\rm M}_{\rm Jup}$}

\begin{document}
	\title{A New Window into Planet Formation and Migration: Refractory-to-Volatile Elemental Ratios in Ultra-hot Jupiters}

	\author[0000-0003-3667-8633]{Joshua D. Lothringer}
	\affiliation{Department of Physics and Astronomy, Johns Hopkins University, Baltimore, MD, USA}
	
	\author[0000-0003-4408-0463]{Zafar Rustamkulov}
	\affiliation{Department of Earth \& Planetary Sciences, Johns Hopkins University, Baltimore, MD, USA}
	
	\author[0000-0001-6050-7645]{David K. Sing}
	\affiliation{Department of Physics and Astronomy, Johns Hopkins University, Baltimore, MD, USA}
	\affiliation{Department of Earth \& Planetary Sciences, Johns Hopkins University, Baltimore, MD, USA}
	
	\author[0000-0002-9308-2353]{Neale P. Gibson}
	\affiliation{School of Physics, Trinity College Dublin, Dublin 2, Ireland}
	
	\author{Jamie Wilson}
	\affiliation{Astrophysics Research Centre, School of Mathematics and Physics, Queens University Belfast, Belfast BT7 1NN, UK}
	
	\author[0000-0001-5761-6779]{Kevin C. Schlaufman}
	\affiliation{Department of Physics and Astronomy, Johns Hopkins University, Baltimore, MD, USA}

	\vspace{0.5\baselineskip}
	\received{November 20th, 2021}
	\revised{March 26th, 2021}
	\accepted{April 14th, 2021}
	\email{jlothri1@jhu.edu}
	
	\begin{abstract}
A primary goal of exoplanet characterization is to use a planet's current composition to understand how that planet formed. For example, the C/O ratio has long been recognized as carrying important information on the chemistry of volatile species. Refractory elements, like Fe, Mg, and Si, are usually not considered in this conversation because they condense into solids like Fe(s) or MgSiO$_3$ and would be removed from the observable, gaseous atmosphere in exoplanets cooler than about 2000~K. However, planets hotter than about 2000~K, called ultra-hot Jupiters (UHJs), are warm enough to largely avoid the condensation of refractory species. In this paper, we explore the insight that the measurement of refractory abundances can provide into a planet's origins. Through refractory-to-volatile elemental abundance ratios, we can estimate a planet's atmospheric rock-to-ice fraction and constrain planet formation and migration scenarios. We first relate a planet's present-day refractory-to-volatile ratio to its rock-to-ice ratio from formation using various compositional models for the rocky and icy components of the protoplanetary disk. We  discuss potential confounding factors like the sequestration of heavy metals in the core and condensation. We then show such a measurement using atmospheric retrievals of the low-resolution UV-IR transmission spectrum of WASP-121b with PETRA, from which we estimate a refractory-to-volatile ratio of 5.0$^{+6.0}_{-2.7}\times$ solar and a rock-to-ice ratio greater than 2/3. This result is consistent with significant atmospheric enrichment by rocky planetismals. Lastly, we discuss the rich future potential for measuring refractory-to-volatile ratios in ultra-hot Jupiters with the arrival of JWST and by combining observations at low- and high-resolution.
		
\end{abstract}
	

\section{Introduction}\label{intro}

Planet formation, migration, and evolution is a complex process that both observation and theory seek to understand in better detail. Because the chemistry of a protoplanetary disk can vary dramatically with orbital distance from the host star, the present-day composition of mature exoplanets may hold the imprint of the planet formation process by keeping information on where in the disk the planet formed and how it may have migrated.

When a giant planet reaches a critical core mass of about 10~$\mearth$ through core accretion, the planet's gravity will be strong enough to begin accreting significant amounts of gas in a runaway growth phase \citep{pollack:1996,lissauer:2007}. Because the core needs to reach a high mass before the dispersal of the gaseous protoplanetary disk, giant planets are often thought to form out beyond the H$_2$O iceline where ice provides an extra reservoir of solid material to help quickly form the protoplanet core. Initially, it was thought that there was not enough mass in either solids or gas inside the H$_2$O iceline to form a giant planet quickly enough to accrete gas. Rather, hot Jupiters are thought to be gas giants that formed beyond the iceline and subsequently migrated to their present-day short periods through interaction with the gaseous protoplanetary disk (Type II migration); \citep{papaloizou:2007,kley:2012} or through a dynamical interaction with a third body, sending the planet to shorter-periods on an eccentric orbit that eventually circularizes \citep{rasio:1996,fabrycky:2007,winter:2020}.

Recently, however, there has been some theoretical exploration of the \textit{in situ} formation of hot Jupiters in the inner portion of the disk, showing that a small number of proto-sub-Jovians may grow quickly enough, perhaps through collisions, to reach the critical core mass around 10~$\mearth$ and initiate runaway gas accretion \citep{boley:2016,batygin:2016}. The inner edge of the orbital distance-mass space has been used as demographical evidence of the \text{in situ} formation of hot Jupiters \citep{bailey:2018}.

After the initial formation stage, additional planetesimals will enrich the planet's envelope. This, combined with potential mixing of the gaseous atmosphere with the planet's primordial rock-ice core will determine the final observable composition of the planet. Thus, when we measure the elemental abundances of an exoplanet, we can begin to trace the composition of a planet's building blocks.

\subsection{Elemental Abundances from Planet Host Stars}

Variation in elemental abundances between exoplanet host stars and non-planet hosting stars can potentially lend insight into both the composition of planets and their formation \citep{bond:2008}. For example, the Sun shows a 10\% depletion in refractory elements when compared to nearby solar twins \citep[e.g.,][]{melendez:2009}. However, rather than these elements being incorporated into planets themselves, \cite{booth:2020} suggest that planets can trap dust exterior to the planets orbit, preventing accretion of this dust onto the host star.

It may also be possible to tie elemental abundances in host stars to exoplanet demographics. For example, a robust relationship exists between the stellar metallicity of a host and the occurrence of giant planets \citep{fischer:2005,buchhave:2014}. This relationship illuminates our understanding of planet formation by pointing to more efficient core accretion in protoplanetary disks with high heavy element content. One might logically take this a step farther and assume metal-rich stars might host metal-rich planets, yet this does not seem to clearly be the case \citep{teske:2019}. However, a potential correlation does exist between the planet's metal content inferred from its bulk density and the stellar volatile-to-refractory elemental abundance, perhaps pointing to efficient heavy element enrichment in planets in disks with more ices, but this relationship is tentative.

One last way that stellar elemental abundances can reveal planet composition is through the pollution of white dwarf atmospheres by orbiting rocky material \citep{zuckerman:2003}. The composition of this rocky material can be measured quite precisely and represent one of the very few ways we might be able to understand the makeup of this rocky material. In general, this material resembles Earth's bulk  composition, mainly containing O, Mg, Si, and Fe with depleted C \citep[e.g.,][]{jura:2014}.

\subsection{Elemental Abundances in the Solar System}

Variation in elemental abundances has been a focus of solar system research for decades, with several instruments built and sent to Jupiter and Saturn to make such measurements \citep[e.g.,][]{niemann:1998,wong:2004,li:2020,grassi:2020}. Combined with remote observations of the solar system giant planets, these efforts have placed constraints of 3.3-5.5, 9.5-10.3, 71-100, and 67-111$\times$ solar on the atmospheric metallicities of Jupiter, Saturn, Neptune, and Uranus  based on the C/H ratio measured from CH$_4$ \citep{karkoschka:2009,karkoschka:2011,sromovsky:2011,fletcher:2009b}. Such measurements have long been used to constrain the location of giant planet formation in the solar system \citep[e.g.,][]{lodders:2004,mousis:2009a,oberg:2019,bosman:2019}. While \textit{in situ} mass spectroscopy can measure even the noble gases in solar system giant planets, refractory elements like Fe, Mg, and Si remain locked away, condensed in the deep interior. Such elements were the first to condense out of the solar nebula and therefore are the first building blocks of the planets. 

Because the chemistry of the a protoplanetary disk depends on location and time \citep{eistrup:2018}, estimates of Jupiter's bulk rock-to-ice ratio range from about 0.5 to 2 depending on disk chemistry and Jupiter's migration history during formation and enrichment \citep{lodders:2003,thiabaud:2014,venturini:2020}. While gravity field measurements from spacecraft like \textit{Juno} can provide constraints on the interior mass distribution \citep{hubbard:1989,wahl:2017}, the composition of this mass is quite degenerate for interior structure models \citep{grasset:2017,ni:2019}. Interior models for Uranus and Neptune indicate rock-to-ice ratios of 2.8-5.3\% and 7.1-28\%, respectively \citep{nettelmann:2013}. This low ratio for the ice giants remains puzzling for formation models to explain, but interior models of such planets may also be inadequate \citep{grasset:2017}. 

The bulk rock-to-ice ratio can be measured for outer solar system moons via their density \citep{wong:2008,encrenaz:2008}, however these densities show significant variation between satellites \citep{johnson:2005}. Evidenced by the decreasing rock-to-ice ratio with distance for the Galilean satellites, the details of the circumplanetary disk likely matter more in moon formation than the protoplanetary disk \citep{stevenson:1986,ronnet:2016}. Nonetheless, the rock-to-ice ratio of $\sim1$ found in Ganymede and Callisto may be representative of the available material that formed Jupiter \citep{mousis:2009a}.

The rock-to-ice ratio has also been measured in cometary material, with 1P/Halley having a ratio of 1 or greater \citep{mcdonnell:1987,mcdonnell:1991,fulle:2000}. Observations of comet 9P/Tempel 1 from the Deep Impact mission suggest a dust-to-ice ratio greater than 1 \citep{kuppers:2005}. Multiple analyses of observations of comet 67P/Churyumov-Gerasimenko from ESA's Rossetta mission measure a refractory-to-ice ratio of 3-7, though measurements from the COmetary Secondary Ion Mass Analyzer instrument imply a lower value \citep{fulle:2017,choukroun:2020}.

Comet formation models that include coagulation of particles and aggregation via the streaming instability suggest high rock-to-ice ratios of up to 3-9 \citep[][]{lorek:2016}. Note that the refractory component in comets can include high amounts of organics and hydrocarbons, and so the comets are not necessarily low in C or O, which we discuss as volatile elements in this work. For example, even with a high refractory-to-ice ratio, comet 67P appears to have a nearly solar Fe/C ratio \citep{fulle:2017}. 

Thus a determination of the rock-to-ice ratio in exoplanets from the direct measurement of volatile and refractory species could provide insight into fundamental questions about the formation of our own solar system.

\subsection{Elemental Abundances Outside the Solar System}

The atmospheres of exoplanets afford us the same opportunity to study elemental enrichment as in the solar system, but for planets light years away. The shear number of exoplanets that we can characterize opens the door to a more comprehensive understanding of the outcomes of planet formation. Historically, only a handful of atmospheric species have been unambiguously characterizable with current instrumentation, namely H$_2$O from the Hubble Space Telescope's (HST) Wide Field Camera 3 (WFC3; \citealt[e.g.,][]{deming:2013,wakeford:2013,kreidberg:2014b,sing:2016}) and the alkali metals, Na and K, from both space \citep[e.g.,][]{charbonneau:2002,nikolov:2014} and the ground \citep[e.g.,][]{wyttenbach:2015,nikolov:2016,casasayas:2018,seidel:2019}. Limited constraints on carbon are possible from \textit{Spitzer} photometry \citep{kreidberg:2015,spake:2020}, but CO and CO$_2$ cannot be distinguished. High-resolution observations of hot Jupiters have detected CO \citep[e.g.,][]{brogi:2014,dekok:2013} and potentially CH$_4$ \citep{guilluy:2019}, in addition to H$_2$O \citep[e.g.,][]{birkby:2017}, but work is still proceeding on how best to obtain and interpret atmospheric constraints from the continuum-normalized high-resolution observations \citep[e.g.,][]{brogi:2017,brogi:2018,pino:2018,gibson:2020,fisher:2020}. Direct imaging of exoplanets has similarly been able to measure H$_2$O, CO, and CH$_4$ \citep[e.g.,][]{konopacky:2013,barman:2015,samland:2017,greenbaum:2018,molliere:2020} with, e.g., C/O constrained to be 0.43 $\pm 0.05$ in $\beta$ Pic b \citep{gravity:2020}.

In the end, most of our determinations of exoplanetary atmospheric elemental abundances have heretofore depended only on measurements of O/H from H$_2$O as well as Na/H and C/H to some degree \citep[e.g.,][]{kreidberg:2015,wakeford:2018,welbanks:2019a}. O/H in particular has been used as a tracer of the bulk metallicity of a planet, yet it is not clear whether O/H ought to be representative of the overall elemental abundances. While the C/O ratio has been well-studied for its sensitivity to the planets formation location \citep[e.g.,][]{oberg:2011,madhusudhan:2012b,mordasini:2016}, both C and O mostly constrain the composition of a planet's ice. To get a full inventory of the rock and ice, we must look at other elemental abundances in addition to C and O.

Recently, ultra-hot Jupiters (i.e., jovian planets with $T_{eq}\gtrsim 2,000$~K) have been recognized as particularly spectrally rich due to the simple fact that their atmospheres are hot enough to largely avoid the condensation of even refractory materials \citep{lothringer:2018b,kitzmann:2018,parmentier:2018,lothringer:2020b}. This allows elements like Mg and Fe to remain gaseous, and thus observable, throughout the atmosphere, unlike cooler planets where those elements will participate in condensation and thus be removed from the gaseous atmosphere. Indeed, a true plethora of atmospheric species have now been detected in several ultra-hot Jupiters. KELT-9b, the hottest ultra-hot Jupiter \citep{gaudi:2017}, was the first planet where multiple atomic metals were detected, including Fe I, Fe II, Mg I, Ti II, Na I, Cr II, Sc II, and Y II, with evidence for 4 other elements in addition to detections of H$\alpha$ and H$\beta$ \citep{yan:2018,hoeijmakers:2018a,hoeijmakers:2019,cauley:2019}. As predicted by \cite{lothringer:2018b}, Fe has now been seen in emission from the dayside spectrum of KELT-9b, indicating a strong temperature inversion \citep{pino:2020}. WASP-121b has a similarly rich transmission spectrum as KELT-9b with detections of Mg I, Mg II, Na I, Ca I, Cr I, Fe I, Fe II, Ni I, and V I along with extended H$\alpha$ absorption \citep[e.g.,][]{sing:2019,hoeijmakers:2020b,ben-yami:2020,bourrier:2020,gibson:2020,merritt:2020,cabot:2020,borsa:2020}. At low-resolution, these metals contribute to increased transit depths at short wavelengths \citep{lothringer:2020b}, with additional features of H$_2$O and VO \citep{evans:2018,evans:2019}. Other characterized ultra-hot Jupiters with spectrally rich transmission and/or thermal emission spectra include WASP-33b, WASP-76b, MASCARA-2b/KELT-20b, and WASP-189b \citep{nugroho:2017,nugroho:2020,nugroho:2020b,fu:2020,casasayas:2019,yan:2020}.

Here, we argue that these very metals, found only in their gaseous form in the atmospheres of ultra-hot planets, can place unique constraints on planet formation and migration through the measurement of the refractory-to-volatile element ratios. In Section~\ref{sec2}, we relate a planet's present-day refractory-to-volatile elemental abundances to the rock-to-ice ratio from the planet's formation and discuss confounding factors to interpreting refractory-to-volatile ratios. In Section~\ref{retrievals}, we show an example of such a measurement using a retrieval of WASP-121b's low-resolution HST transmission spectrum. In Section~\ref{discuss}, we discuss the future of measuring refractory-to-volatile ratios with JWST and high-dispersion spectroscopy before concluding in Section~\ref{conclude}.

\section{Tying Elemental Abundances to Rock and Ice}\label{sec2}

To first order, planets are made up of rock, ice, and gas. Rock in this case means some combination of metals (FeS, Al$_2$O$_3$), silicates (e.g., MgSiO$_3$, Mg$_2$SiO$_4$), and any other refractory material. Ices, on the other hand, represent the volatile condensates present only at low temperatures, such as H$_2$O, CO, and CO$_2$ ice. The gas then makes up the rest of the planet and contains everything that has not yet condensed. The composition of these components will vary as a function of temperature in the disk \citep[e.g.,][]{oberg:2011}, which itself will evolve. Chemical evolution in the disk also appears to be important with volatile abundances changing with ionization level \citep{eistrup:2018,notsu:2020}.

\begin{figure}[ht!]
	\centering
	\textbf{Rock Composition}\par\medskip
	\underline{Species by Mass Fraction} \par\medskip
	
	\includegraphics[width=3.5in]{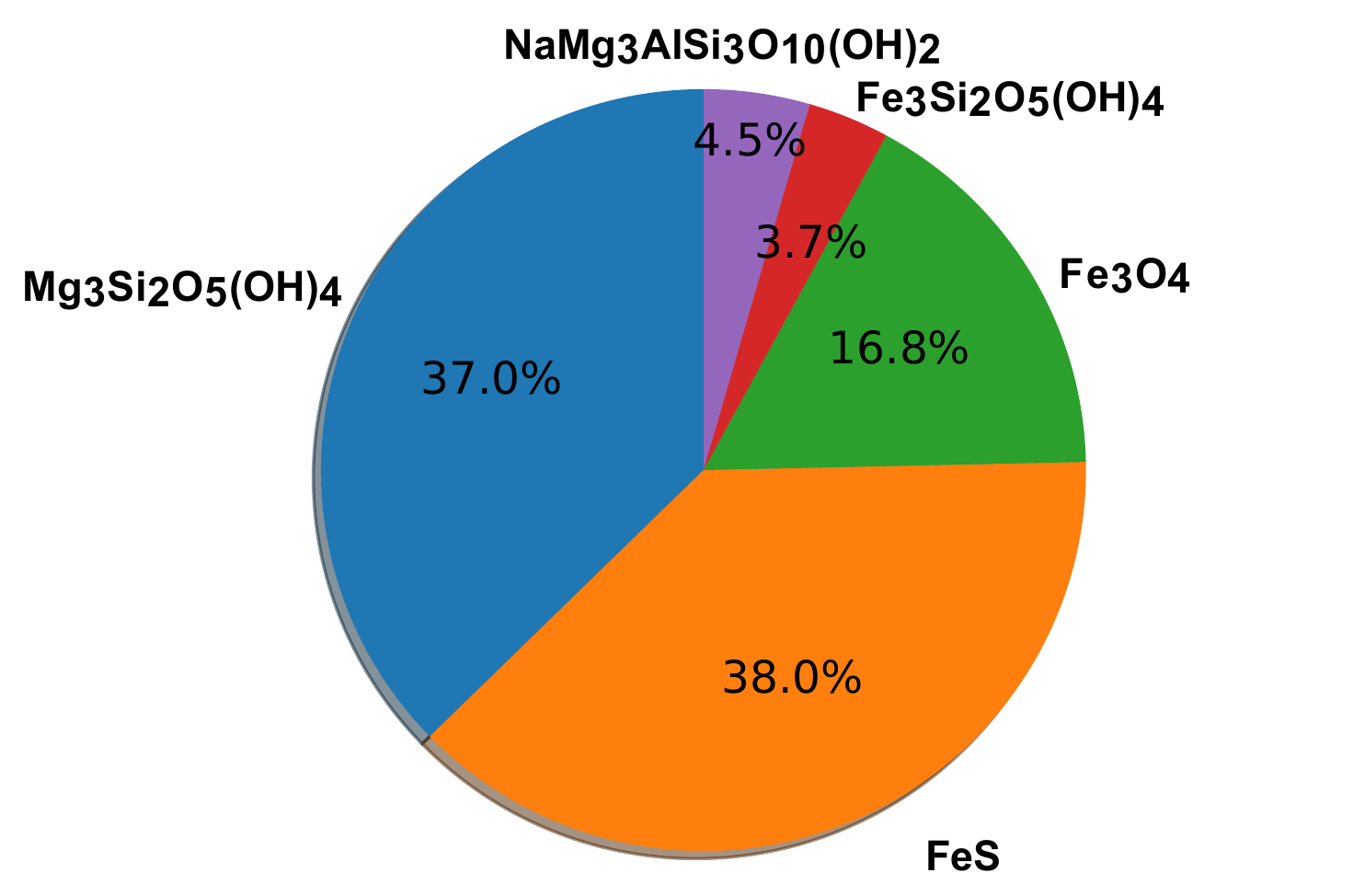} 
	\underline{Elements by Mixing Ratio} \par\medskip
	\includegraphics[width=3.5in]{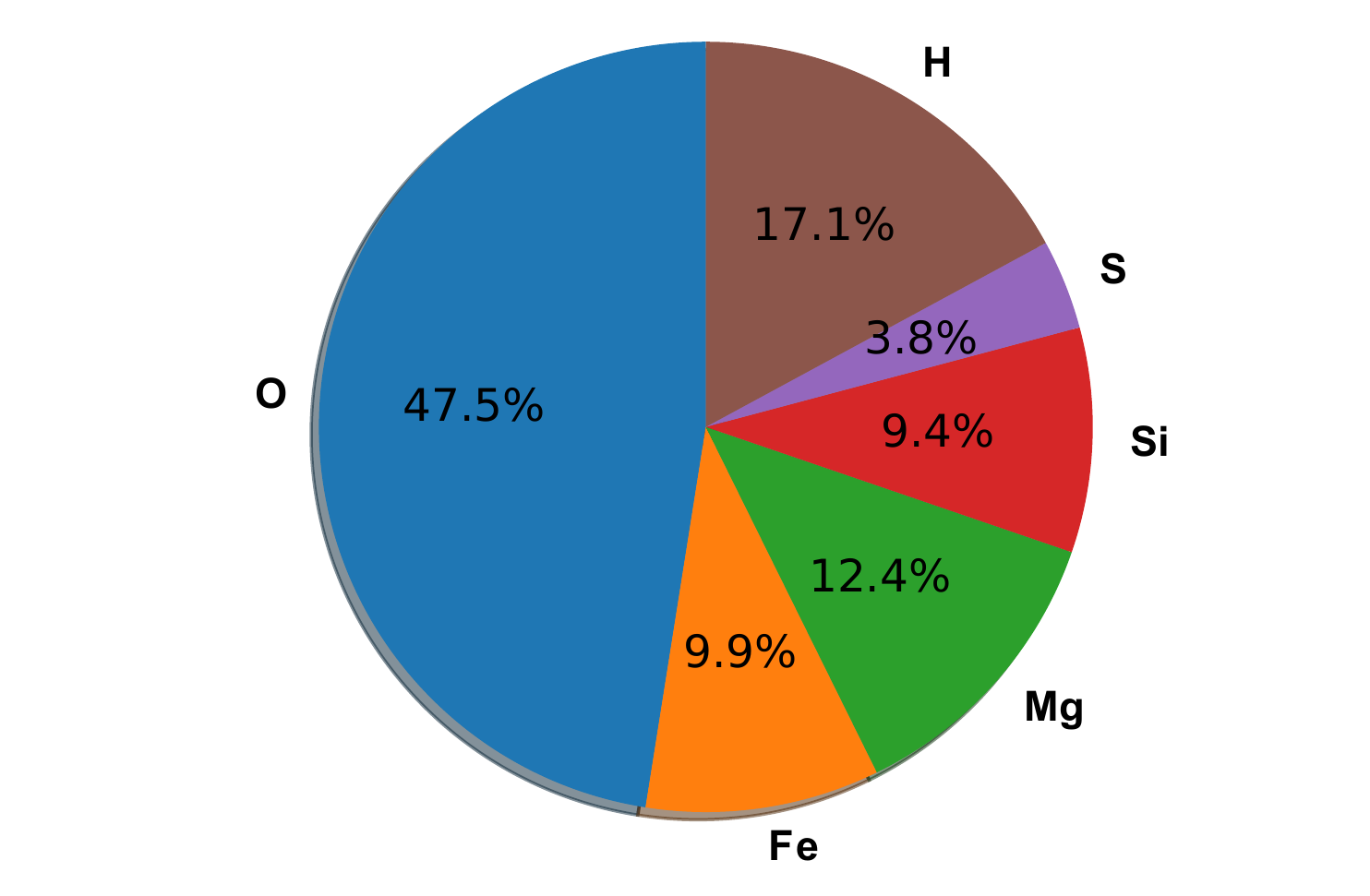} 
	\caption{Top: Mass fraction of rock species making up our adopted rock composition, using GGchem \citep{woitke:2018}. We neglect species with mass fractions $<$1\%. Bottom: Elemental mixing ratios for the rock composition. While no refractory carbon is predicted by GGchem, we also include a scenario where 55\% of the primordial carbon is refractory. \label{fig:rock}}
\end{figure}

\begin{figure}[ht!]
	\centering
	\textbf{Ice Composition}\par\medskip
	\underline{Species by Mass Fraction} \par\medskip
	\includegraphics[width=3.5in]{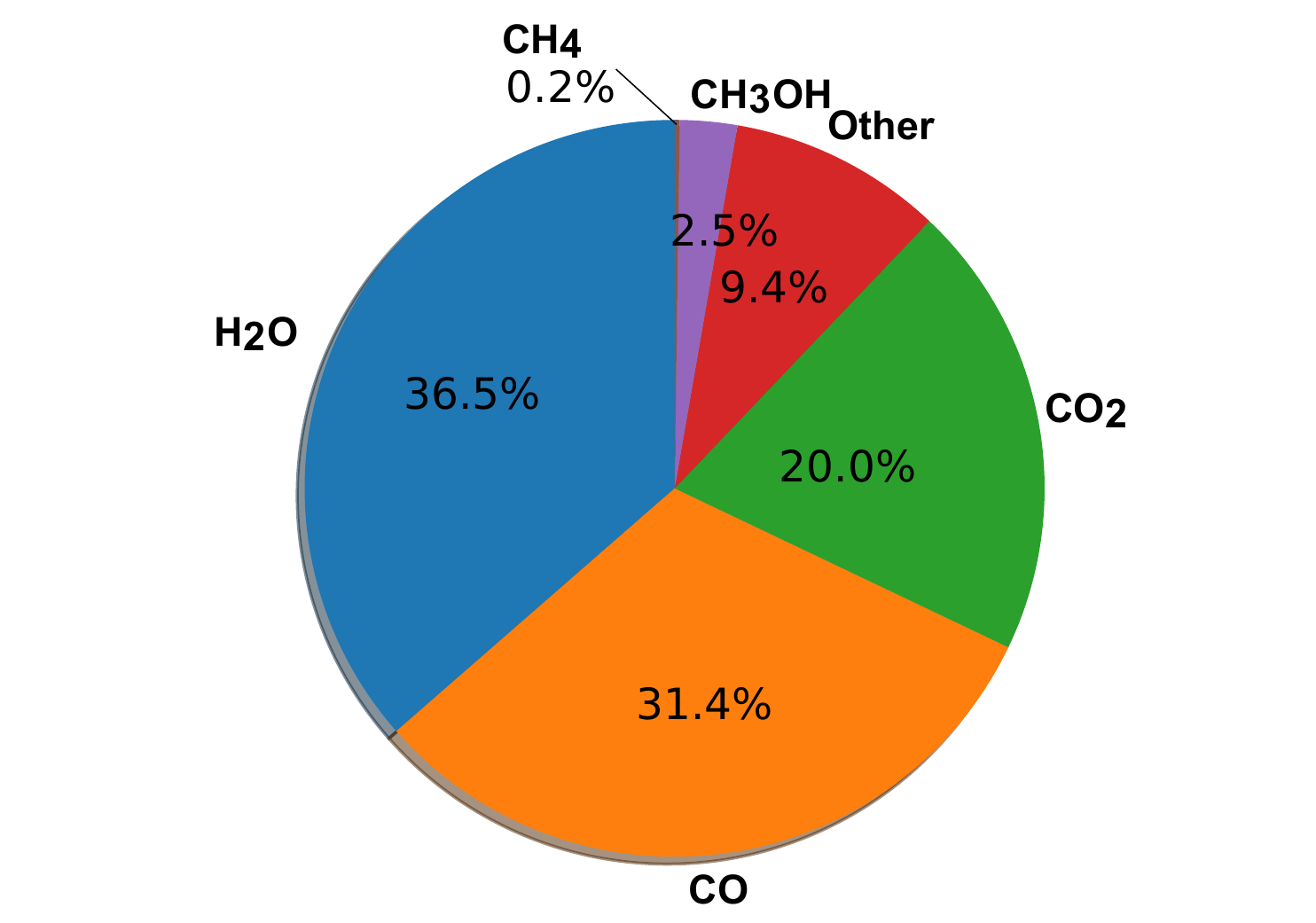} 
	\underline{Elements by Mixing Ratio} \par\medskip
	\includegraphics[width=3.5in]{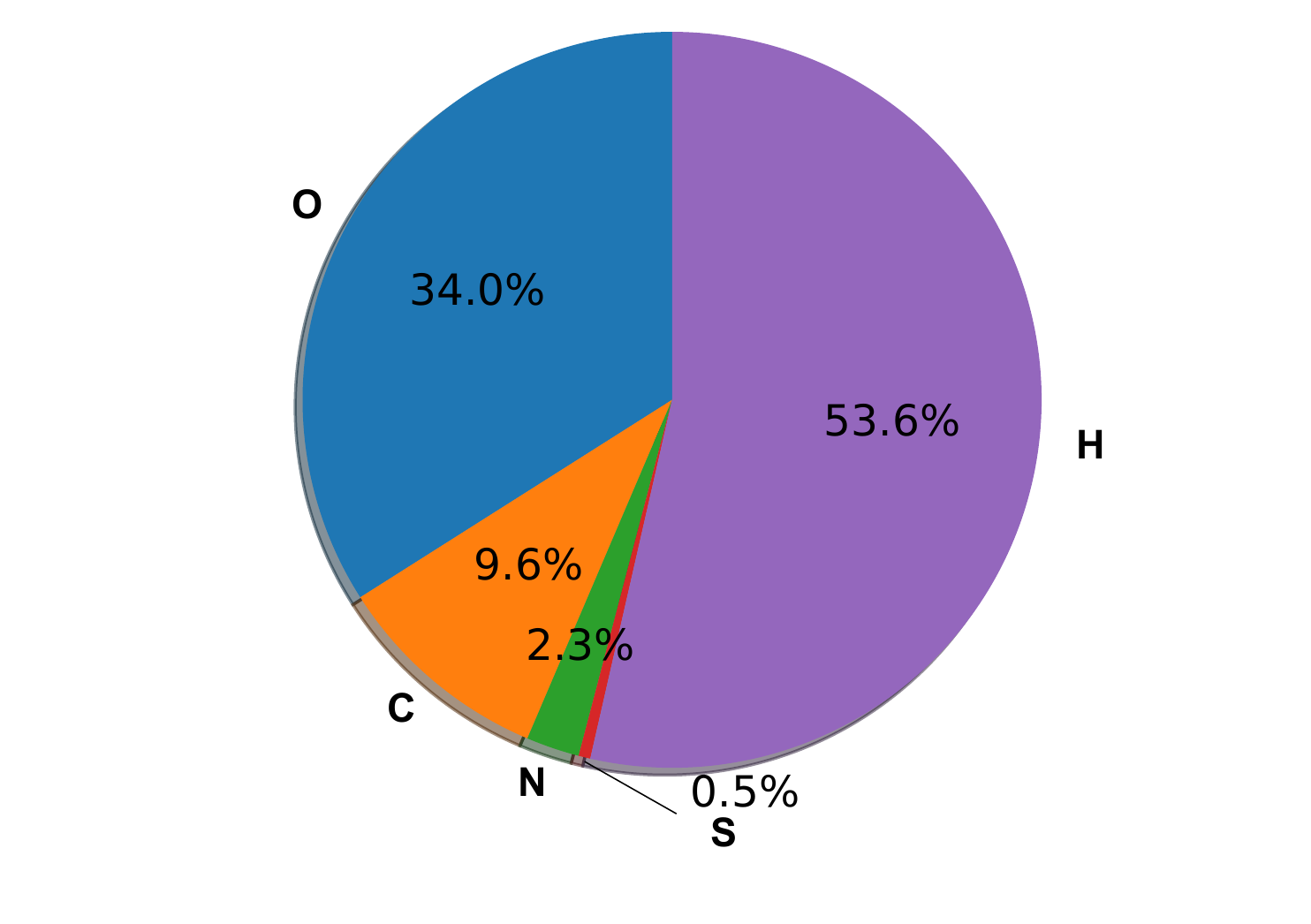} 
	\caption{Top: Mass fraction of ice species making up our adopted ice composition from \cite{mousis:2009a}. ``Other" accounts for all other ice species that do not include C or O (e.g., NH$_3$ and H$_2$S). Bottom: Elemental mixing ratios for the ice composition. \label{fig:ice}}
\end{figure}

Here, we explore how a planet's present-day elemental abundances will vary if the planet is enriched with different proportions of rock and ice. We must first adopt a composition for these components. To determine the composition of the refractory species in the disk we assume chemical equilibrium, similar to past chemical studies of rocky exoplanet formation \citep{bond:2010a,bond:2010,johnson:2012,elser:2012,thiabaud:2014}. We use GGchem, a publicly available chemical equilibrium solver which is accurate down to 100~K, below which numerical convergence becomes difficult \citep{woitke:2018}. GGchem minimizes the Gibbs free energy and numerical convergence becomes difficult below 100~K. The version we use considers 552 gaseous species and about 241 condensed (i.e., solid and liquid) species with elements up to tungsten.

We take the rocky component of the planetesimals to be the condensed species at T = 200~K at 0.1 mbar assuming solar abundances, where we expect all refractories to have condensed with only volatile species, like H$_2$O, remaining gaseous. Our assumption of solar composition can be improved for individual systems given measurements of their stellar composition. The intrinsic variation in stellar elemental abundances can be large enough to affect the inferred composition of planets formed in such systems, especially if the stellar C/O ratio is high \cite[e.g.,][]{bond:2010,johnson:2012}. Formation at different temperatures and oxidation states may also be explored, as this will affect the refractory elemental ratios if formation occurs at short orbital periods \citep{pekmezci:2020}. For the scope of this paper, however, we present the case of solar elemental abundances where all refractory elements have condensed.

Figure~\ref{fig:rock} shows the species that make up the rock, mostly composed of magnesium-iron silicates and iron oxides. GGchem predicts phyllosilicates (e.g., Mg$_3$Si$_2$O$_5$(OH)$_4$, or serpentine) to be the dominate silicate at 200~K, with \cite{thi:2020} arguing that phyllosilicates can form in disk conditions. Figure~\ref{fig:rock} also shows the resulting elemental composition of the rock, which generally agree with other definitions of rocky compositions \citep{hubbard:1989,pollack:1994}. As expected, all refractory elements are entirely condensed such that all Fe, Mg, Si, etc. are contained in the rock. Much of the oxygen remains in volatile species like H$_2$O, CO, and CO$_2$, but a significant fraction of total oxygen in the disk, about 30\%, is found in refractory species (also see \citealt{pollack:1994}).

GGchem does not predict any refractory carbon at solar C/O elemental abundances, despite considering C (graphite), SiC, and TiC. One reason this may be is that GGchem does not consider organic material in its chemical equilibrium solution. This importantly means that the rock component of the planetesimals will have a C/O of 0. This scenario is similar to other models without refractory carbon and/or with carbon-depletion \cite[e.g.,][]{mordasini:2016,cridland:2019}. Interior to a so-called `tar-line' \citep[][]{lodders:2004}, refractory carbon can be destroyed in the inner solar system by evaporation \citep{nakano:2003} and reactions at the grain-gas interface \citep{lee:2010:carbon}. Refractory carbon can also be removed through photo-chemical processes \citep{alata:2014,anderson:2017}, which can release enough carbon to raise the gas-phase C/O ratio above 1.5 \cite{bosman:2021}.

There is some observational evidence that this may be true in protoplanetary disks. The carbon abundance in asteroids is deficient by a factor of $\gtrsim$10 \citep{wasson:1988,bergin:2015} and by about 10$^4$ in Earth's crust \citep[e.g.,][]{wang:2018} compared to refractory elements. However, carbon does not appear to be depleted in comets \citep{geiss:1987,min:2005}. Thus, there appears to be a gradient in the carbon abundance as a function of formation location. The polluted atmospheres of white dwarfs by extrasolar rocky material also points to low amounts of refractory carbon \citep{jura:2006}. This is different from what was originally assumed in planet formation models \citep[e.g.,][]{pollack:1994}, which was based on the composition of interstellar gas, where carbon organics can form from UV photolysis of ice \citep{jenniskens:1993}, and on evidence that some comets contain carbon in the form of organics \citep{jessberger:1988}.

We also consider a case where the rock does have refractory carbon. We put 55\% of the primordial carbon into the rock composition, similar to \cite{pollack:1994}. In this case, carbon elemental ratios behave more similarly to oxygen elemental ratios with respect to the rock-to-ice ratio since both elements make up a portion of the rock and ice.

We take our ice composition from \cite{mousis:2009a}, which calculated the condensation sequence of ice in the outer solar system. We use the case that assumes 10\% clathration efficiency as the disk cools. Clathration is the process whereby some species in the gas phase become trapped in the condensing H$_2$O ice. A 10\% clathration efficiency thus assumes that only some clathration cages are filled and we choose it as a middle ground between no clathration and total clathration. In the end, the effect of clathration on the resulting elemental composition of the ice is minor, at most 5-10\% for the major C and O bearing ices. The mass fraction of ice species and the subsequent elemental composition is shown in Figure~\ref{fig:ice}. The C/O ratio in the ices is low (0.28) compared to solar (0.55), underscoring the difficulty in forming giant planets with high C/O ratios if the final heavy metal content is driven by planetesimal enrichment. This agrees with similar calculations in \cite{marboeuf:2014b} and \cite{espinoza:2017}. 

The ice composition can also vary greatly with location in the disk, so a planet may form where H$_2$O ice is the only available ice species. These possible scenarios can be explored by measuring multiple refractory and volatile species (e.g., O and C with JWST). Future studies may explore the possible range of ice and rock compositions in a more self-consistent fashion, as well \citep[e.g.,][]{eistrup:2016,mordasini:2016,cridland:2019}.

\cite{thiabaud:2015b} and \cite{mordasini:2016} showed that for Jovian planets less than 2-8\mjup, the enrichment of a planet's composition during its formation and migration is dominated by the accretion of additional planetesimals, rather than accretion of additional gas, whose composition has been altered from condensation \citep[e.g.,][]{oberg:2011}. If this is the case, then the elemental ratios observable in exoplanet atmospheres are determined by the rock-to-ice ratio of these planetesimals \citep[e.g.,][]{espinoza:2017}. So, from the adopted rock and ice compositions, we then calculate different elemental ratios as a function of the mass fraction of rock (i.e., varying the amount of rock in the planetesimals).

In what follows, we will use the variable $R$ to refer to the collection of the refractory elements. $R$/O will then represent the refractory-to-oxygen ratio and [$R$/O] will represent the logarithmic refractory-to-oxygen ratio relative to solar. The individual elemental mixing ratios (i.e., mole fraction) for a given rock-to-ice ratio from our composition of the rock and ice by mass are given by:

\begin{equation}
    x_i = x_{i,ice} (1-Z_\textrm{rock}) + x_{i,rock} (Z_\textrm{rock}) 
\end{equation}

and

\begin{equation}
    x_{i,\textrm{ice or rock}} = \frac{w_i/M_i}{\sum\limits_{j} w_j/M_j}
\end{equation}

\noindent where $x_i$ is the mixing ratio of element $i$ with $x_{i,ice}$ being the mixing ratio of element $i$ in the ice and $x_{i,rock}$ being the mixing ratio in the rock. $Z_\textrm{rock}$ is then the total mass fraction of rock in the planetesimals, which equals $(1-Z_\textrm{ice})$. $w_i$ is the mass fraction of element $i$ in the ice or rock and $M_i$ is the molar mass of element $i$ in the ice or rock.

When discussing proportions of rock and ice, we use rock-to-ice ratio $Z_\textrm{rock}/Z_\textrm{ice}$, where $Z_\textrm{rock}$ is the total mass fraction of rock and $Z_\textrm{ice}$ is the total mass fraction of ice. In plots, we will also use the mass fraction of rock in the solids, or
\begin{equation}
Z_\textrm{s,rock}= Z_\textrm{rock}/(Z_\textrm{ice}+Z_\textrm{rock})
\end{equation}
to avoid large numbers nearing completely rocky compositions. For example, a rock-to-ice ratio of 2 would correspond with a rock mass fraction of 2/3.

\begin{figure}[t]
	\centering
	\includegraphics[width=3.5in]{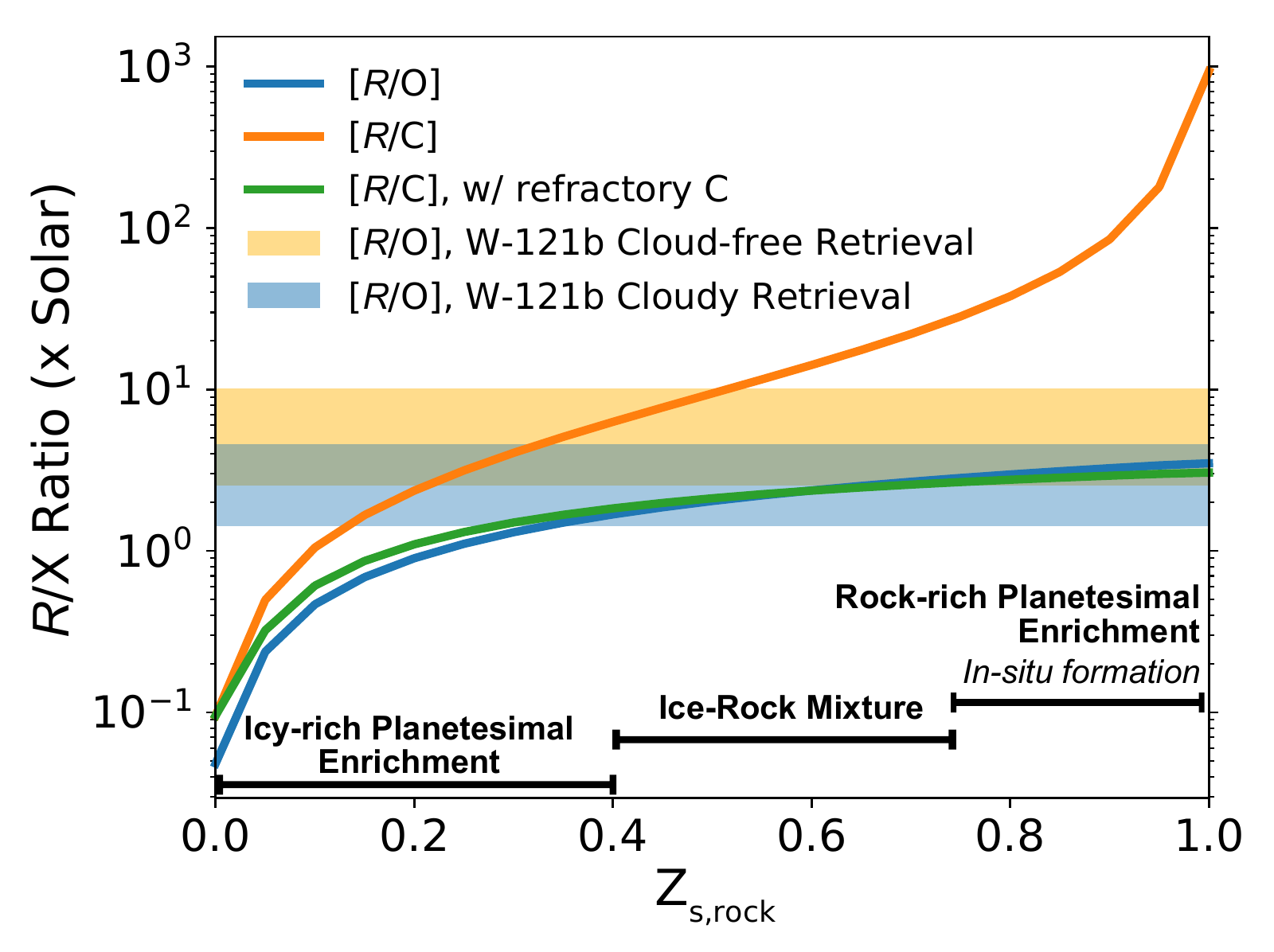} 
	\caption{The ratio of refractory elements, $R$, to the oxygen and carbon abundances relative to solar as a function of mass fraction of rock in the planetesimals that enrich the planet's atmosphere (with ice making up the remaining fraction of planetesimal mass). Rock elemental composition is calculated using GGchem \citep{woitke:2018}, while the ice elemental composition is taken from \cite{mousis:2009a}. We show the 1-$\sigma$ range of our measurement of WASP-121b from the PETRA retrievals of the low-resolution HST/STIS and WFC3 transmission spectrum from \cite{evans:2018} described in Section~\ref{retrievals}. \label{fig:zrx}}
\end{figure}

Figure~\ref{fig:zrx} shows [$R$/O] and [$R$/C], the $R$/O and $R$/C ratios relative to their solar values, as a function of $Z_\textrm{s,rock}$. When stellar elemental abundances are available, they should be used to define $[R]$ rather than solar values, but the intrinsic variation of stellar $R_*$ is smaller than the precision we are presently likely to reach in measuring the planet's $R$/O \citep{nissen:2014,teske:2019}. Nonetheless, it will be helpful for future exploration of exoplanet elemental ratios if the corresponding stellar values are also known, e.g., via high-resolution spectroscopy.

Importantly, the relationship between the elemental ratios and the mass fraction of rock is monotonic so an elemental ratio can be associated with a single mass fraction of rock. This simple calculation depends only on the assumed composition of rock and ice and some conclusions are straightforward (e.g., a high [$R$/O] and [$R$/C] imply rock-rich planetesimals). Very high [$R$/C] ratios are possible in situations where there is no refractory carbon allowing a tight constraint on the mass fraction of rock. When some amount of carbon is found in refractory phases, [$R$/C] behaves more like [$R$/O] which becomes more linear towards high mass fractions of rock. 

\subsection{Tying the Rock-to-Ice Ratio to Formation Location}

Going one step further, we can estimate how $Z_\textrm{s,rock}$ might vary throughout the disk, allowing us to constrain the formation location and/or determine the migration history of a given planet. To first order, a high mass fraction of rock can imply an \textit{in situ} formation and/or enrichment, since rock-dominated planetesimals will only be expected in the inner disk, inside the H$_2$O iceline \citep{boley:2016,batygin:2016}. On the other hand, a very low mass fraction of rock would imply formation and/or enrichment in the outer portion of the disk, followed \textit{afterwards} by migration through a relatively clear disk. Measurements of a moderate rock-to-ice ratio would indicate a mixture of both rocky and icy planetesimals, which might occur as a planet migrates from the outer disk \citep{fogg:2007,venturini:2020}. We discuss caveats to these simple interpretations in Sections~\ref{sec:core}-\ref{sec:frac}.

In agreement with this intuition, planet formation models predict that the mass fraction of rock in planetesimals, $Z_\textrm{s,rock}$, gets smaller at farther distances from the host star. \cite{marboeuf:2014}, for example, show that $Z_\textrm{s,rock}$ will begin to decrease from 100\% at the H$_2$O iceline, as we would expect. Interior to that point, ices are unstable due to the high temperature. $Z_\textrm{s,rock}$ will gradually decrease as more ices begin to condense, like CO and CO$_2$, until a minimum of about 50\% is reached around 10-30 AU \citep{marboeuf:2014}. 

While beyond the scope of this paper, future studies may model the ice composition consistently with the distance from the host star to determine the direct relationship between distance from the host star and the present-day refractory-to-volatile elemental ratios similar to \cite[e.g.,][]{mordasini:2016}. Since the inner disk will have both a high surface density \citep[][]{weidenschilling:1977} and a high rock-to-ice ratio, planetesimals accreted in the inner disk may be most important for determining $Z_\textrm{s,rock}$ in short period planets. However, large icy bodies (i.e., comets) can survive interior to the H$_2$O iceline, serving to enrich the atmosphere of giant planets even when they are already at their short orbital periods \citep[][]{mandt:2015}. We also note that the location of the icelines changes throughout the disks lifetime so $Z_\textrm{s,rock}$ is both a function of time and distance from the host star. Future exploration of these questions can help us better interpret any rock-to-ice ratios we measure in exoplanet atmospheres.

While we focus here on core accretion formation, which is likely most relevant for planets forming closer than several tens of AU from their host star, formation through gravitational instability \citep[][]{boss:1997} may be important at larger orbital distances \citep[][]{boley:2009}. \cite{madhusudhan:2014} suggest that sub-solar O/H ratios may be indicative of hot Jupiter formation by gravitational instability followed disk-free migration to present-day short periods. The measurement of low refractory abundances would help confirm such a scenario, if possible.

\subsection{Elemental Enrichment versus Bulk Enrichment}

A difference in the rock-to-ice fraction, and thus the refractory-to-volatile ratio, will cause the individual elemental abundances to diverge from the ``bulk" metallicity. A planet's composition will be dramatically different if its atmosphere is enriched entirely by rock compared to if it is enriched entirely by ice, even for the same bulk metallicity. Similar uncertainties exist with respect to the C/O ratio \citep{welbanks:2019}. Thus, when we measure elemental ratios like O/H, C/H, and Na/H, it is not clear \textit{a priori} to what degree those ratios represent the bulk composition. This will be especially important in interpreting mass-metallicity relationships to understand how the bulk composition of a planet varies with mass \citep[e.g.,][]{kreidberg:2015,thorngren:2016,wakeford:2017b,welbanks:2019,spake:2020}.

\begin{figure*}[ht!]
	\centering
	\includegraphics[width=3.5in]{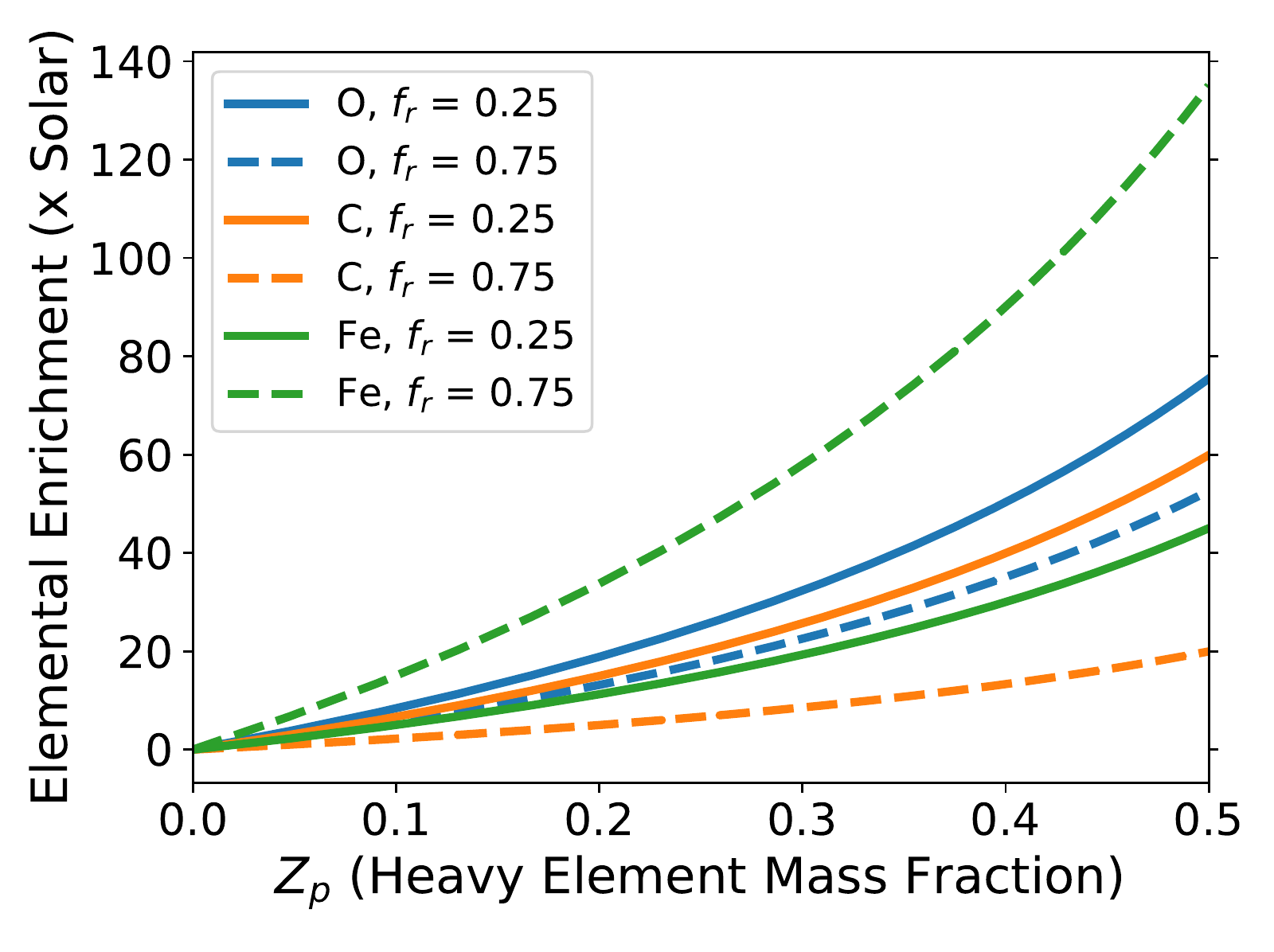} 
	\includegraphics[width=3.5in]{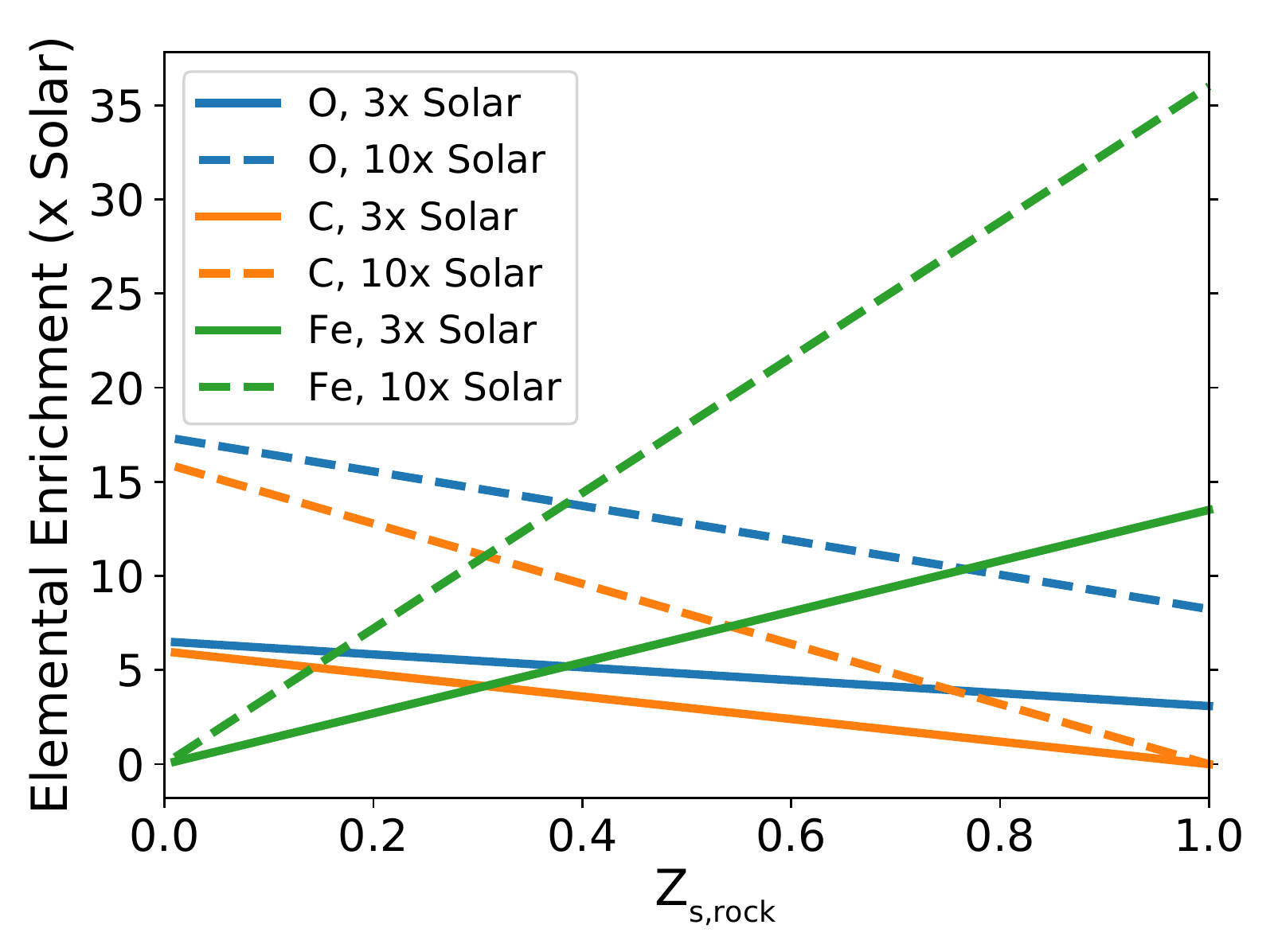} 
	\caption{Left: Elemental enrichment (relative to solar abundance) of O, C, and Fe as a function of total heavy element mass fraction (rock + ice), $Z_\textrm{p}$ for two rock mass fractions, 0.25 and 0.75. Right: Elemental enrichment (relative to solar abundance) of O, C, and Fe as a function of rock mass fraction, $Z_\textrm{s,rock}$, for two different bulk metallicities (3$\times$ and 10$\times$ solar). For clarity, we do not include here the case with refractory carbon. \label{fig:enrich}}
\end{figure*}

Figure~\ref{fig:enrich} shows how the enrichment above solar of individual elements O, C, and Fe varies with respect to both the heavy element mass fraction (i.e., metallicity by mass) and the rock mass fraction, $Z_\textrm{s,rock}$. For a Jupiter-mass exoplanet, a heavy metal mass fraction of 0.2 roughly corresponds with a bulk metallicity of 10$\times$ solar if enrichment was uniform amongst heavy elements. However, for a high rock mass fraction like 0.75 (or rock-to-ice ratio of 3), a heavy metal mass fraction of 0.2 corresponds with an enrichment of Fe of nearly 40$\times$ solar. Similarly, for a low rock mass fraction of 0.25 (or rock-to-ice ratio of 1/3), the O enrichment with a heavy metal mass fraction of 0.2 would be about 20$\times$ solar. Thus our inference of bulk metallicity from individual elemental abundances depends on an understanding of the rock-to-ice ratio and on our assumption of the rock and ice composition itself. 

\subsection{Confining Elements in the Core}\label{sec:core}

If little refractory or icy material capable of enriching the atmospheric envelope is accreted after runaway gas accretion begins \citep{pollack:1996,ikoma:2000,helled:2014b,paardekooper:2004,levison:2010,bitsch:2018}, the elemental abundances of the atmosphere are therefore determined primarily by mixing with the deep interior and enrichment from the gas during accretion. The degree to which giant planets are well-mixed and/or their cores are eroded throughout their evolution is an open area of research, but a portion of a giant planets bulk heavy metal content may be confined in the planet's interior by a lack of convection due to compositional gradients \citep{chabrier:2007,leconte:2012,vazan:2015,vazan:2016,wahl:2017,moll:2017,vazan:2018,stevenson:2020}.

Thus, the degree to which the measured atmospheric elemental enrichments (e.g., C/H) are representative of the bulk elemental enrichment will be difficult to know \textit{a priori}. However, if the abundances in Jupiter are any indication, we may expect the elemental ratios themselves to be preserved. The abundances of noble gases Ar, Kr, and Xe are all essentially equal to about 2.6 $\times$ solar, despite Xe being 3$\times$ more massive than Ar, indicating the elements are not significantly fractionated by mass \citep{mahaffy:2000}. Recent calculations have also shown that rock and ice should remain mixed throughout the interior, especially for highly irradiated planets \citep{vazan:2020}. Additionally, the menagerie of atomic metal detections in ultra-hot Jupiter atmospheres straightforwardly demonstrates that not all heavy elements can be trapped in the interior and that enrichment from planetesimals is important.

When combined with bulk metallicity estimates from a planet's density \citep[e.g.,][]{thorngren:2016,baraffe:2008}, a precise and comprehensive accounting of a planet's atmospheric metal content through the measurement of refractory element abundances will provide insight into the fraction of heavy elements in the interior \citep{thorngren:2019a}. For example, the high atmospheric metallicity of WASP-39b measured from H$_2$O \citep{wakeford:2018} may imply a well-mixed interior given its bulk density.

\subsection{Condensation of Heavy Elements}\label{sec:condense}

Ultra-hot Jupiters will not necessarily be cloud-free throughout their entire atmosphere. GCMs show that even for planets like WASP-121b with T$_{\textrm{eq}}$=2400~K, materials like Al$_2$O$_3$, CaTiO$_3$, MgSiO$_3$, and Fe can condense at high latitudes, on the nightside, and on the western/morning terminator \citep{wakeford:2017a,parmentier:2018}. Indeed, \cite{ehrenreich:2020} showed evidence of Fe absorption on the cloud-free evening terminator, followed by nightside condensation and a lack of Fe absorption on the morning terminator. Similarly, \cite{hoeijmakers:2020b} proposed that the lack of Ti (and TiO \citep{evans:2018,merritt:2020}) combined with the significant detection of the presumably less abundant V, could be caused by condensation of Ti-bearing species. 

While this complicates the interpretation of refractory abundances in ultra-hot Jupiters, we can at least confidently interpret such abundances as lower limits. Additionally, if absorption from one limb can be isolated, as in \cite{ehrenreich:2020}, we may be able to account for condensation on the other limb. This may also emphasize the hottest ultra-hot Jupiters as the best targets for measuring refractory abundances.

\subsection{Mass Fractionation from Atmospheric Evolution}\label{sec:frac}

While ultra-hot Jupiters are likely to experience some amount of atmospheric escape \citep{yelle:2004,garcia-munoz:2007,murray-clay:2009,owen:2012}, this escape is not expected to dramatically alter the planet's mass or composition for planets more massive than Saturn \citep{mordasini:2016,fossati:2018}. Even the hottest known Jovian exoplanet, KELT-9b, has likely only lost about 1.5-2\% of its total mass over the course of its 300-500 Myr lifetime \citep{wyttenbach:2020}. Some heavy metals have been observed to escape from ultra-hot Jupiters \citep{fossati:2010,sing:2019}, also pointing to heavy elements becoming coupled to the outflow \citep{koskinen:2012a}, which would prevent or reduce any mass fractionation from atmospheric escape \citep{hunten:1987}.

The same cannot be said for small planets. With the discovery of planets like LTT-9779b, one of the first ultra-hot Neptunes, we can begin to consider measuring refractory element abundances in sub-Jovian exoplanets \citep{jenkins:2020}. LTT-9779b exists in the middle of the hot Neptune desert, where gaseous sub-Jovian planets are rare, likely due to their inability to hold onto large low-mean molecular weight atmospheres \citep{matsako:2016,mazeh:2016,owen:2018}. Thus, planets in the hot Neptune desert are thought to currently be undergoing significant mass-loss. 

This hydrodynamic atmospheric loss can be mass-dependent, in that lighter elements may escape more easily compared to heavy elements \citep{hunten:1987}. This mass-fractionation could dramatically alter the composition of these planets and increase the refractory-to-volatile ratio. On the other hand, if heavy atoms and ions become entrained in the outflow \citep{koskinen:2012a}, the effect may not be as dramatic. For planets that are still hot enough for refractory elements to remain uncondensed, the details of the atmospheric evolution may be investigated by measuring metal atomic absorption at short wavelength and volatile molecular absorption in the IR as we discuss here.

\section{Refractory-to-Volatile Measurement in WASP-121b}\label{retrievals}

\subsection{PETRA}

WASP-121b has one of the most precise optical and NIR transmission spectra for an ultra-hot Jupiter and is currently our best chance to understand the detailed composition of an ultra-hot Jupiter. We use the PETRA retrieval framework \citep{lothringer:2020a} to retrieve the refractory-to-volatile ratio of WASP-121b \citep{delrez:2016} using the low-resolution transmission spectra from \cite{evans:2018}. \cite{lothringer:2020b} showed that the low-resolution NUV-blue transmission spectrum of several ultra-hot Jupiters can be explained by the presence of atomic metal opacities without the need for any disequilibrium process. In particular, atomic metal lines from a variety of neutral and ion species, including (but not limited to) Fe, Ni, Mg, V, Ca, and Cr create substantial absorption shortward of 0.45 microns, effectively creating a blueward slope. Importantly, these species have been detected in the atmosphere of WASP-121b at high-spectral resolution \citep{sing:2019,hoeijmakers:2020b,ben-yami:2020,bourrier:2020,gibson:2020,merritt:2020,cabot:2020,borsa:2020}. With a detection of H$_2$O with HST/WFC3, WASP-121b's transmission spectrum has spectral features from both refractory and volatile species, allowing for the first direct estimate of a planet's refractory-to-volatile ratio, and therefore its rock-to-ice ratio.

PETRA is a retrieval framework built around the PHOENIX atmosphere model. In particular PETRA takes advantage of PHOENIX's extensive opacity database, which contains molecular line lists from over 130 species and atomic opacity up to uranium. The abundance of these species are parameterized in two ways. First, the H$_2$O, TiO, and VO abundances are treated as free parameters and can vary independently between a log$_{10}$ volume mixing ratio (VMR) of -12 and -1. Because there are too many refractory opacity sources to parameterize individually as free parameters, the rest of the atmosphere is then treated in chemical equilibrium, with elemental abundances scaling with the free parameter [Fe/H], which represents $R$. We point out that this general [Fe/H] parameter could be biased if the refractory elemental ratios in WASP-121b were highly non-solar. Future high-SNR low- and high-resolution datasets can measure the individual elemental abundances, but for now we vary all refractory species together. We also note that SH is not included in our retrieval because any SH lines present will likely be masked by atomic metals present \citep[][]{hoeijmakers:2020b}. 

We used the temperature-profile parameterization of \cite{parmentier:2014} as implemented in \cite{line:2013}. In the end, however, the observations do not constrain the temperature well with the median profile virtually isothermal. PETRA currently uses a differential evolution Markov Chain (DEMC) as its statistical framework \citep{terbraak:2006,terbraak:2008}, with which we compare 3,000 iterations of 20 chains. 

We ran retrievals with and without opacity from clouds. In our cloudy retrieval, we also include homogenous haze and cloud parameters as defined in \cite{macdonald:2017}. Testing the inclusion of haze is important because the rise in transit depth at short wavelengths can potentially be fit by a scattering slope. We also tested the inclusion of various opacity sources by running a number of retrievals without different major opacity sources (H$_2$O, TiO, VO, and atomic metal lines). We then compared their Bayesian Information Criteria (BIC) to indicate whether the inclusion of such opacity sources were justified.

\subsection{Retrieval Results}

 \begin{table*}[t] 
		\centering  
		\caption{Retrieval Results Summary}
		\label{tab:models} 
		\begin{tabular}{ccccccc}
			\hline    
			& Scenario & $N_\textrm{params}$ & Max. ln($\mathcal{L}$) & $\chi^2_{\nu}$ (K) & $\Delta$BIC & [R/O] \\
			\hline 
			\hline
			Cloud-free & Fiducial & 12 & 623.96 & 1.47 & N/A & 0.70$^{+0.34}_{-0.33}$ \\
			& No H$_2$O & 11 & 611.17 & 1.76 & -21.10 & N/A\\
			& No VO & 11 & 611.74 & 1.77 & -19.96 & 0.9$^{+0.39}_{-0.36}$\\
			& No TiO & 11 & 623.13 & 1.47 & 2.82 & 0.59$^{+0.29}_{-0.25}$ \\
			& No atomic lines\tablenotemark{\footnotesize $\dagger$} & 12 & 621.33 & 1.54 & -5.26 &1.08$^{+0.15}_{-0.17}$\\ 
			& H$_2$O, TiO, VO only & 11 & 615.40 & 1.67 & -12.64 & N/A\\ 
			\hline 
			Cloud+Haze & Fiducial (with aerosols) & 15 & 624.21 & 1.52 &-12.93 & 0.39$^{+0.25}_{-0.29}$\\
			 & No H$_2$O & 14 & 612.57 & 1.82 &-31.73& N/A \\
			& No VO & 14 & 623.83  & 1.81 &-29.41 &0.74$^{+0.31}_{-0.27}$\\
			& No TiO &14 & 617.38 & 1.69 &-9.21 &0.47$^{+0.20}_{-0.21}$ \\
			& No atomic lines\tablenotemark{\footnotesize $\dagger$} & 15 & 617.27  & 1.71 &-26.81& -1.29$^{+1.00}_{-0.78}$\\
			& H$_2$O, TiO, VO only &14 & 621.15 & 1.59 & -14.57  & N/A\\
			\hline
			Flat line & -- & 1 & 591.38 & 2.03 & -15.9 & N/A \\
			\hline

		\end{tabular}
		\tablecomments{$N_\textrm{obs}$ = 88. \tablenotemark{\footnotesize $\dagger$}Atomic metal lines are not included, but lines from metal-bearing molecules (e.g., SiO, metal hydrides) are included. 
		}
	\end{table*}

Table~\ref{tab:models} lists the goodness-of-fit metrics for the various retrieval scenarios computed.  We reach a $\chi^2_\nu$ of 1.47 and 1.52 for the cloud-free and cloudy retrievals, respectively. This is comparable to the fits with the ATMO retrieval from \cite{evans:2018} that achieved $\chi^2_\nu$=1.5. One main difference between the PETRA and ATMO retrieval is the inclusion of 13 data points shortward of 0.47~\microns, which were not fit by ATMO. 
	
Overall, the fiducial cloud-free retrieval is preferred by the BIC calculation, with the exception of the inclusion of TiO opacity, which is not required to fit the data. Conversely, H$_2$O, VO, and opacity from metals (both in atomic and molecular form) are all required to provide the best fit to the observations. While the retrieval that includes clouds marginally fits better than the cloud-free scenario, the cost of the additional free parameters is not justified via the BIC (a $\Delta$BIC$>$2 is positive evidence for the model, while $\Delta$BIC$>$6 is considered strong evidence \citep{kass:1995}). We also note that our fiducial model is preferred over a simple flat line, which is strongly disfavored despite a 49.25 BIC advantage from only having a single free parameter.

\begin{figure*}[ht]
	\centering
	\includegraphics[width=6.5in]{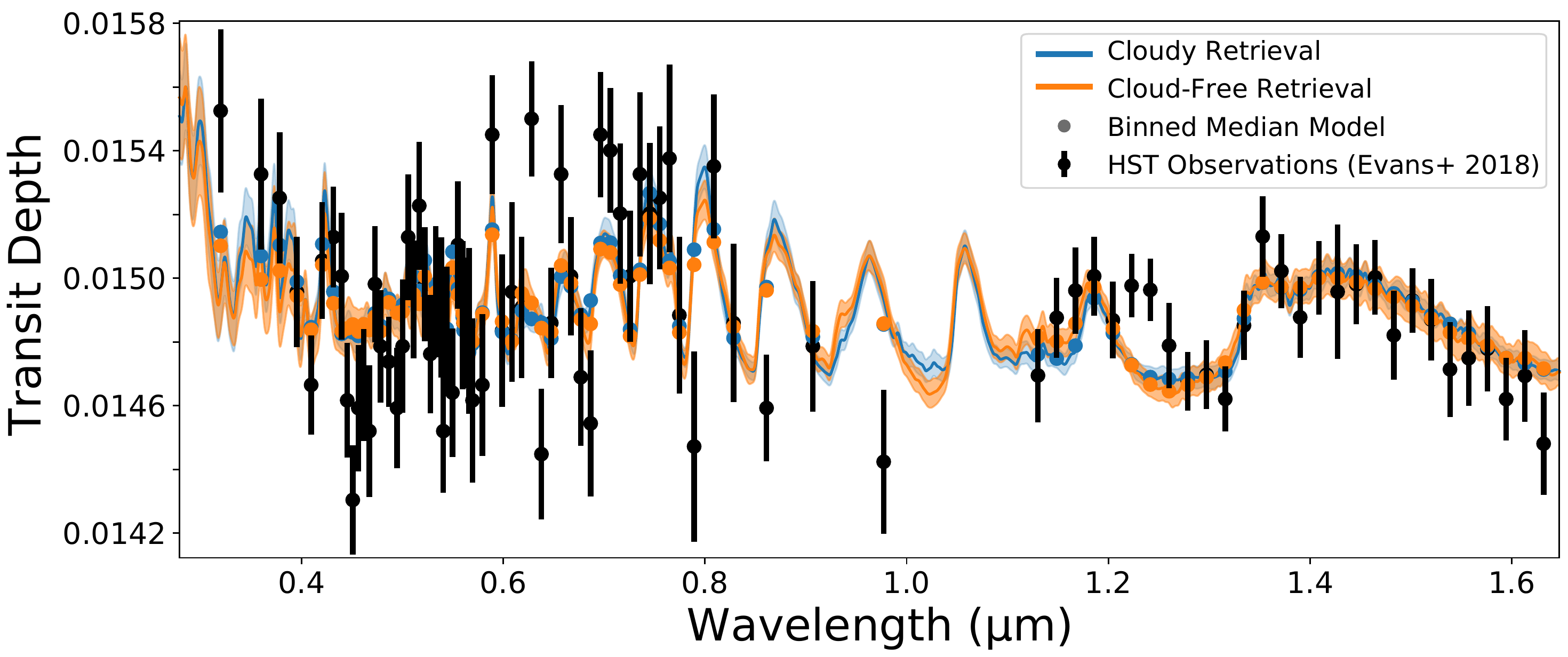} 
	\caption{Retrieved fit to the HST/STIS and WFC3 transmission spectrum of WASP-121b for both a retrieval with and without clouds (blue and orange, respectively). \label{fig:w121}}
\end{figure*}

\begin{figure*}[ht]
	\centering
	\includegraphics[width=7.0in]{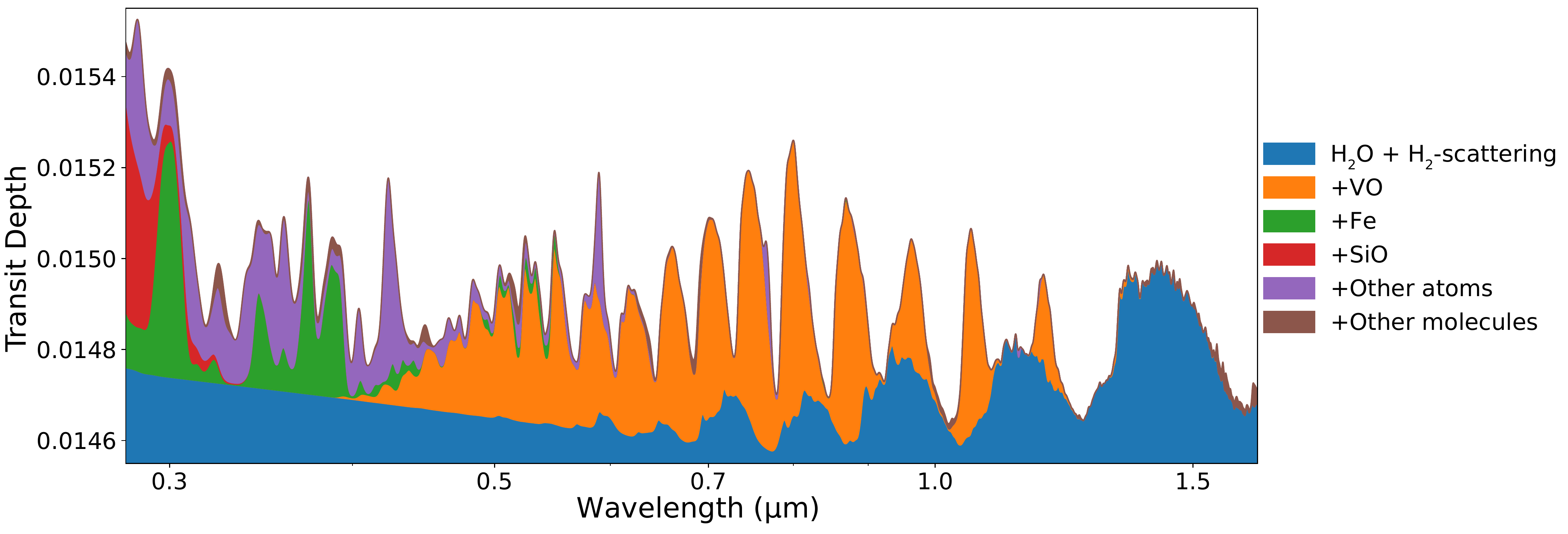} 
	\caption{The contribution of various opacity sources to the best-fit spectrum of the cloud-free retrieval of WASP-121b. Opacity sources are added one after another. \label{fig:w121_contribution}}
\end{figure*}

Figure~\ref{fig:w121} shows the median retrieved spectrum for the fiducial cloud-free and cloudy retrievals compared with the observations. 
Figure~\ref{fig:w121_contribution} demonstrates the contribution of various opacity sources to the best fit spectrum from the cloud-free retrieval. Metal opacity from both atoms and SiO dominate the contribution to the transit depths $<0.5$~\microns{}. This spectral region is where most of the constraint on [Fe/H] comes from in the retrieval. VO is the main opacity source between 0.5 and 1.0~\microns{}, so this region does less to constrain [Fe/H] or H$_2$O, and therefore [$R$/O].

\begin{figure*}[h!]
	\centering
	\includegraphics[width=6.5in]{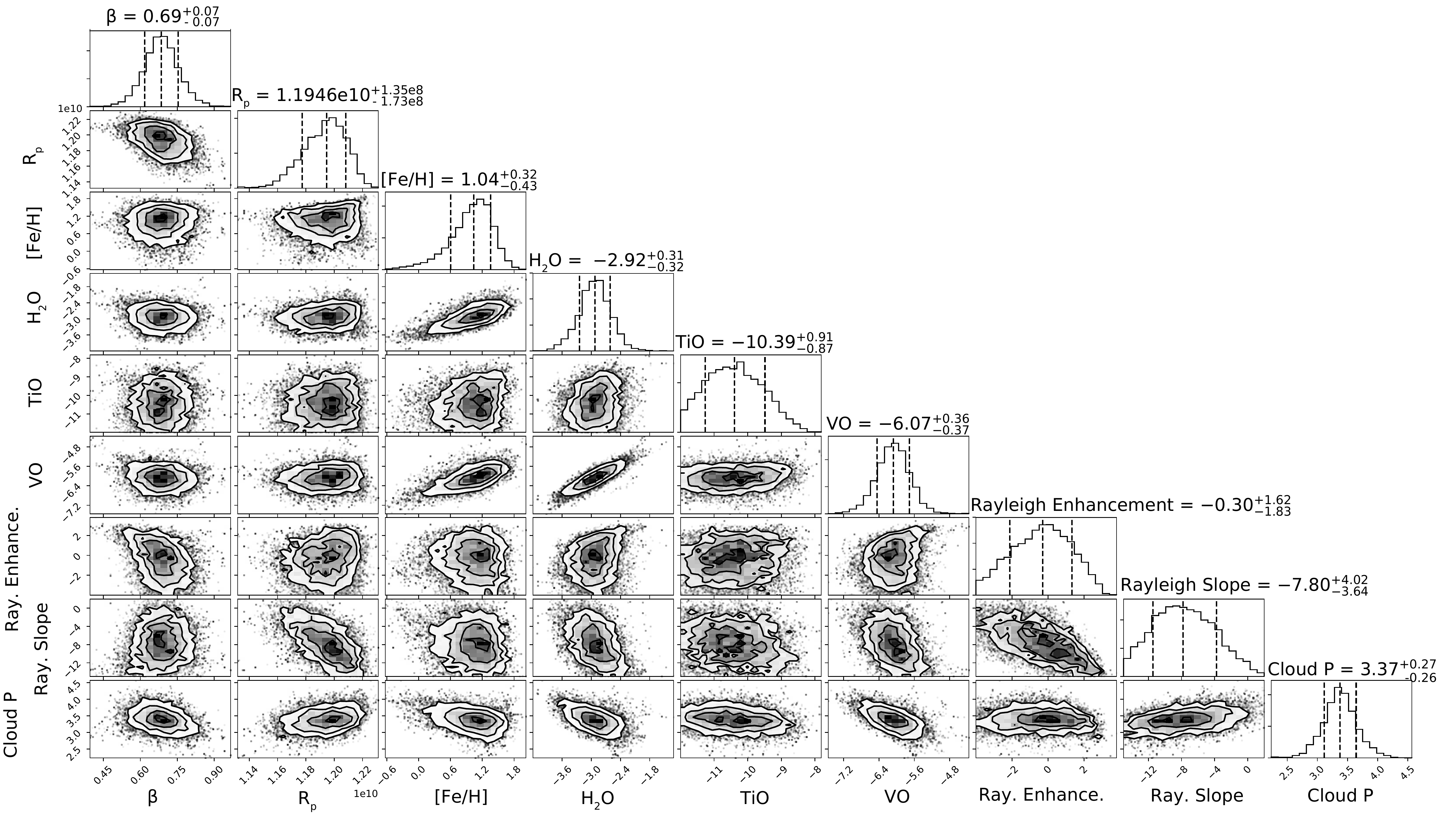} 
	\caption{Retrieved posterior distributions for the retrieval including clouds for WASP-121b using the HST/STIS and WFC3 transmission spectrum from \cite{evans:2018}. \label{fig:corner}}
\end{figure*}

\begin{figure*}[h!]
	\centering
	\includegraphics[width=5.0in]{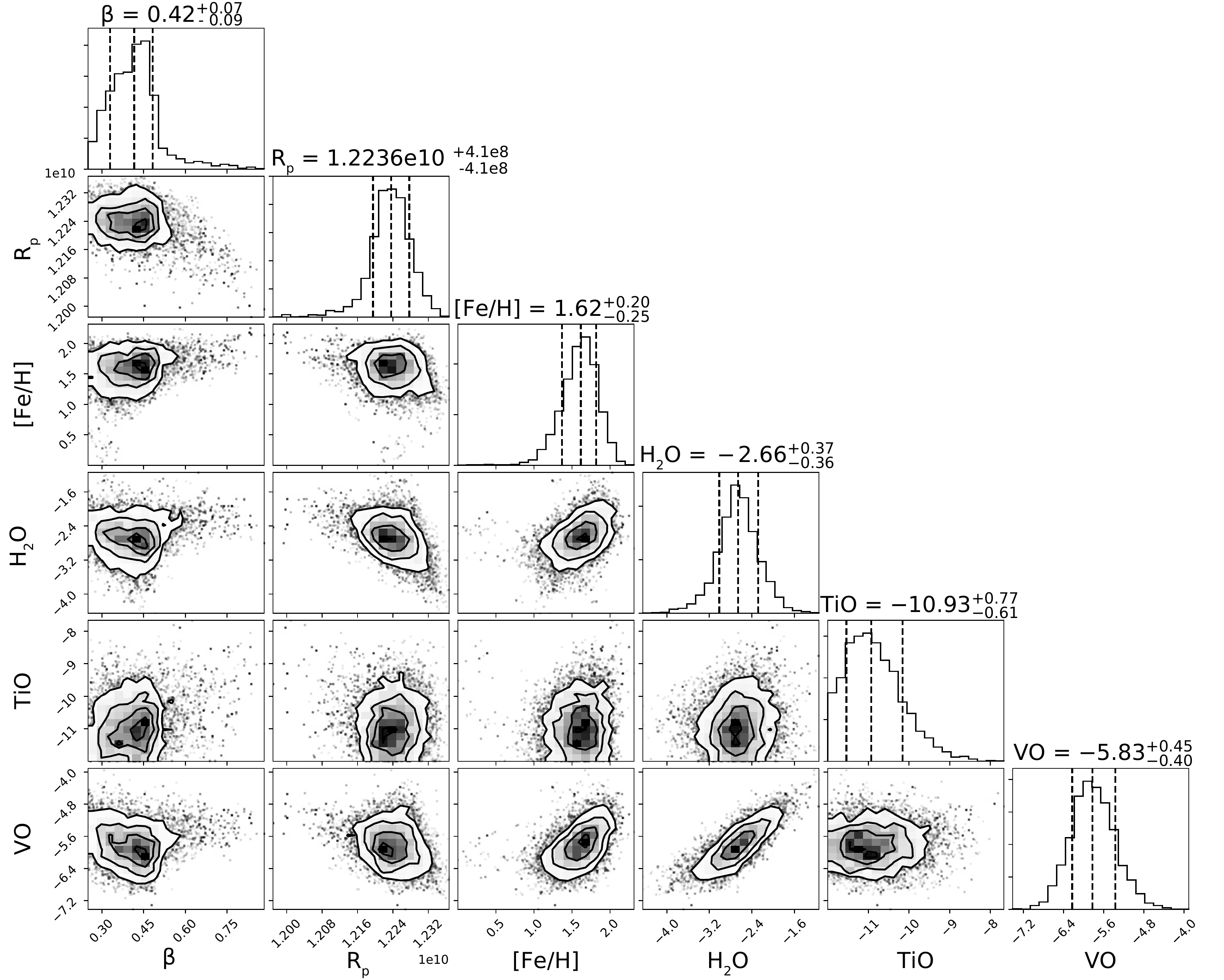} 
	\caption{Same as Figure~\ref{fig:corner}, but for the cloud-free retrieval. \label{fig:corner_cloudfree}}
\end{figure*}

Figure~\ref{fig:corner} and \ref{fig:corner_cloudfree} present the posterior distributions of the fitted parameters for the cloudy and cloud-free retrieval, respectively. Since the median temperature structure is very isothermal, most of the related parameters are not well constrained, thus we only show $\beta$, which controls the average temperature. Our constraints on the atmospheric temperature are $T~=~1630\pm-165$~K for the cloudy retrieval and $T~=~991^{+165}_{-212}$~K for the cloud-free retrieval. As suggested by Table~\ref{tab:models}, opacity from hazes are not necessary to fit the observations, resulting in unconstrained haze parameters. The cloudy retrieval does include a cloud top pressure around 2.3 mbar, which will contain some refractory material. Therefore our measurement of the refractory abundance may be regarded more as a lower limit. While the chemical abundances are correlated with this cloud top pressure, the ratio between the abundances is not. This allows us to put tighter constraints on such abundance ratios than on any individual abundance. While the cloudy retrieval does not fit the observation significantly better than the cloud-free retrieval (and is therefore not justified via the BIC), the retrieved temperature and overall metallicity of the cloudy retrieval is more in line with our theoretical expectations. 

\begin{figure*}[ht]
	\centering
	\includegraphics[width=2.75in]{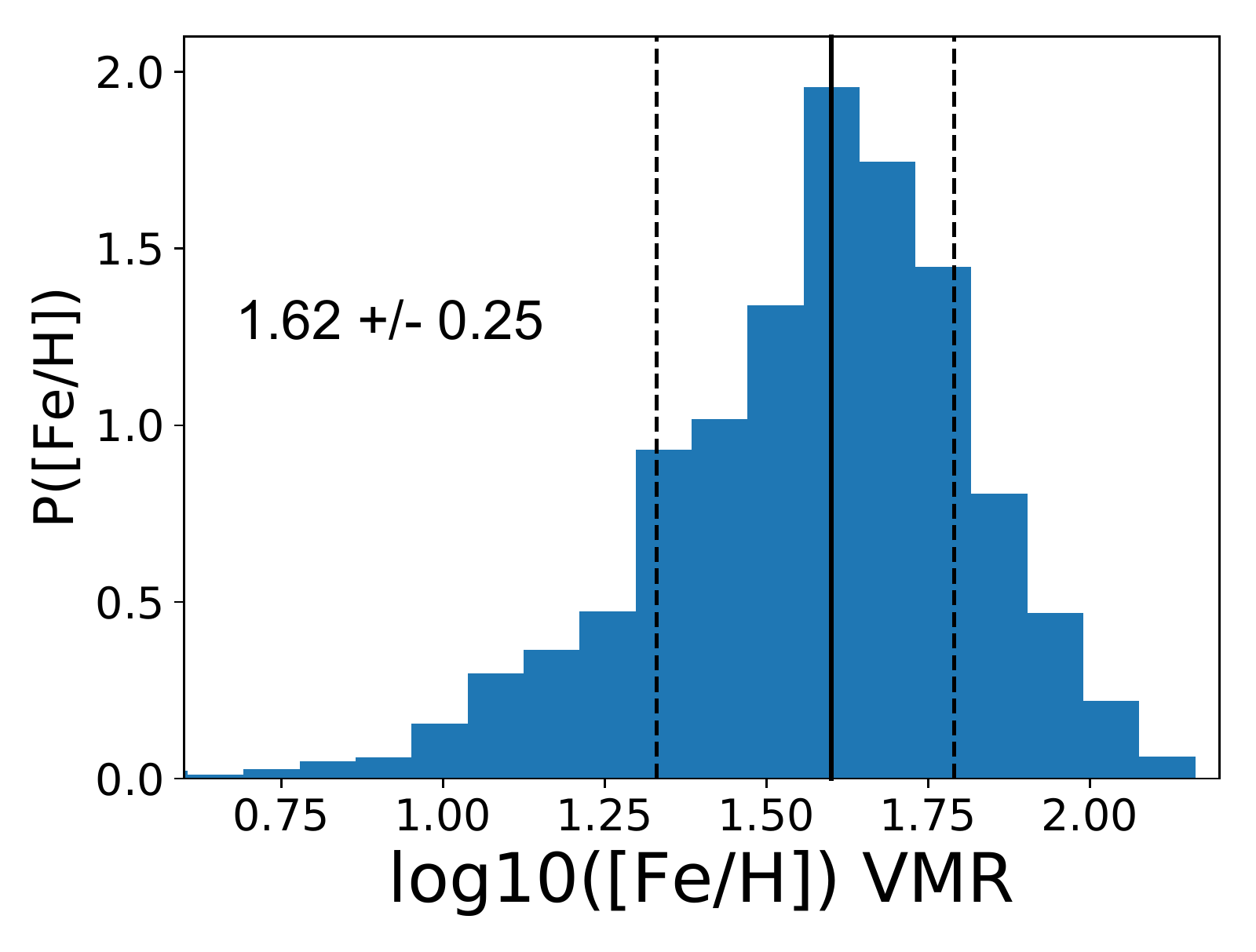} 
	\includegraphics[width=2.75in]{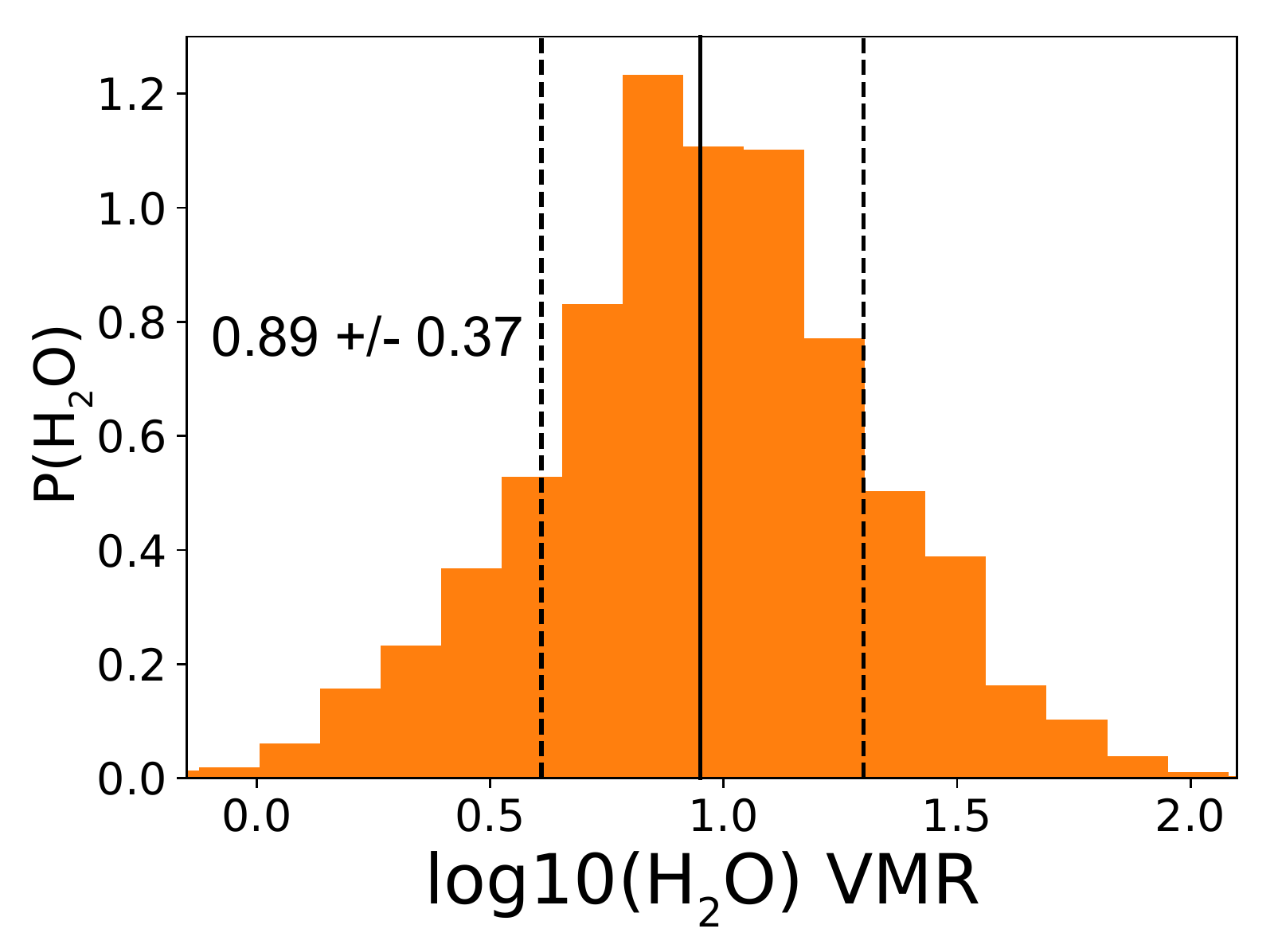} 
	\includegraphics[width=2.75in]{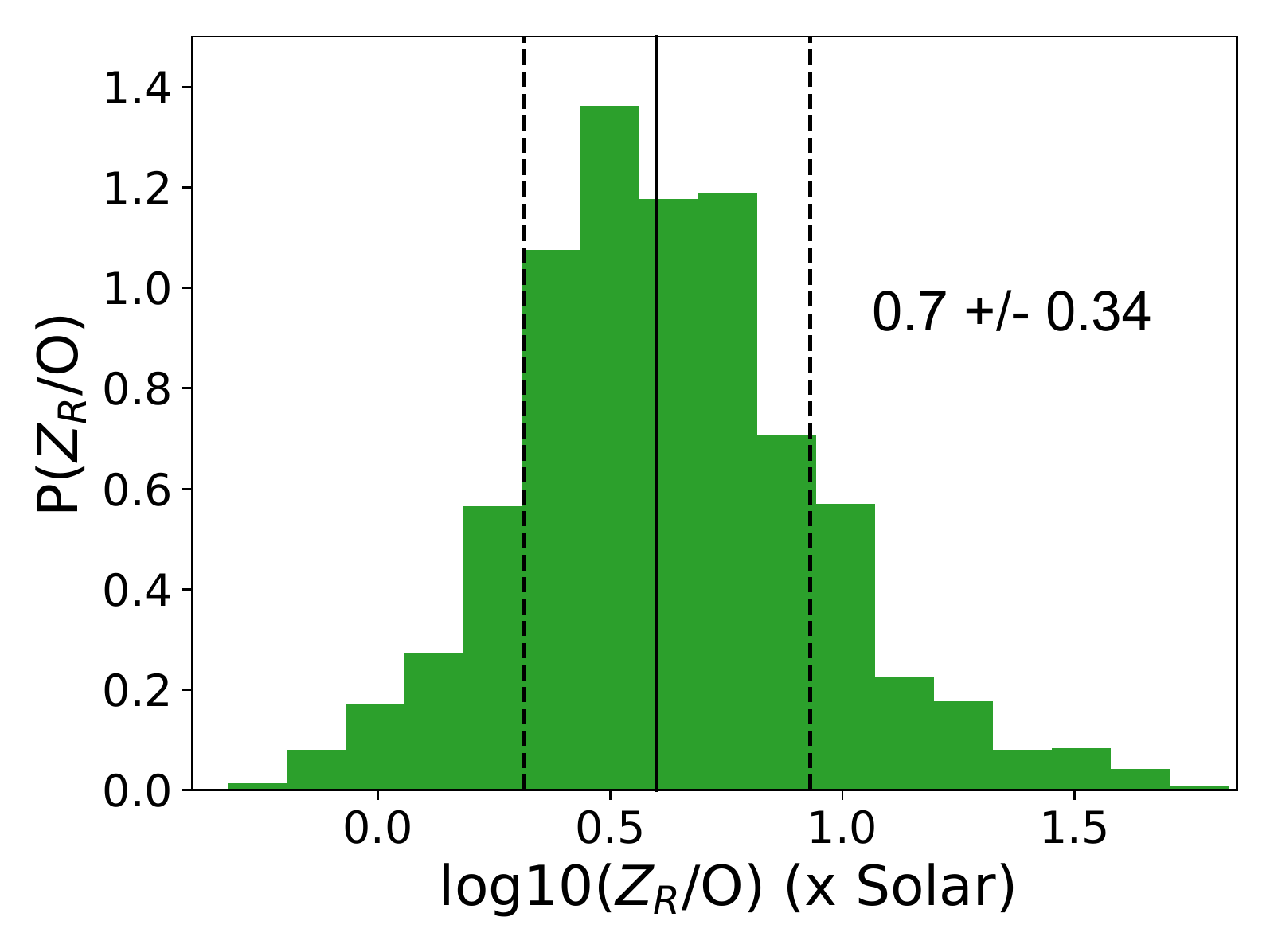} 
	\caption{Retrieved constraints from the cloud-free PETRA retrieval on the chemical abundances, relative to solar, using the HST/STIS and WFC3 transmission spectrum of WASP-121b. \label{fig:w121abund}}
\end{figure*}

Figure~\ref{fig:w121abund} shows the retrieved constraints on the chemical abundances for the cloud-free retrieval in more detail. In what follows, we compare the retrieved abundance ratios to solar since stellar elemental ratios for WASP-121 are not immediately available, though this would improve future analyses. We obtain an H$_2$O abundance that is only somewhat super-solar, at a log$_{10}$(VMR) of -2.66$^{+0.37}_{-0.36}$. A solar metallicity would imply an H$_2$O log$_{10}$(VMR) of about -3.55, constraining [O/H] between 3.3 and 17.8$\times$ solar. The refractory species, as parameterized by [Fe/H], are also super-solar at 1.62$^{+0.20}_{-0.25}$ or between 23.4 and 66 $\times$ solar. Combining the two posteriors, our estimate of [$R$/O] is thus 0.70$^{+0.34}_{-0.33}$ or 5.0$^{+6.0}_{-2.7}$$\times$ solar. Again, because the H$_2$O and [Fe/H] parameters are correlated, we are able to measure the ratio of the two values to a higher precision than either parameter alone. This measurement is consistent with the solar abundance ratio at about 2-$\sigma$. We also note that the ratio of [Fe/H] and [H$_2$O], [$R$/O], remained consistent with our cloudy retrieval at [$R$/O] = 0.39$^{+0.25}_{-0.29}$. This result also seems to rule out a very icy-rich planetesimal enrichment of WASP-121b, pointing to an ice-rock mixture or a rock-rich planetesimal enrichment scenario.

Table~\ref{tab:models} further demonstrates that the retrieved [$R$/O] value remains largely consistent across the different retrieval scenarios. One exception is the retrieval with clouds and hazes where atomic lines are not included but molecular metal opacity, like SiO, is included. The retrieved [$R$/O] is sub-solar with larger uncertainties than other scenarios. In this scenario, strong clouds and hazes are required to fit the blue-ward slope at short-wavelengths, which drives the H$_2$O abundances to very high values ($\sim 100\times$ solar).

We also note that the TiO and VO abundances retrieved in the cloudy case (log$_{10}$(TiO$_{VMR}$) $< -10.35^{+0.89}_{-0.93}$ and log$_{10}$(VO$_{VMR}$) = -6.08$^{+0.35}_{-0.43}$) agree to about 1-2$\sigma$ with with the results of \cite{evans:2018} from the ATMO retrieval (log$_{10}$(TiO$_{VMR}$) $< -10.4^{+1.0}_{-0.9}$ and log$_{10}$(VO$_{VMR}$) = -6.6$^{+0.2}_{-0.3}$). There is a somewhat larger disagreement in the VO abundances retrieved in the clear case (log$_{10}$(VO$_{VMR}$) = -5.81$^{+0.45}_{-0.40}$). Some of this difference could reasonably be ascribed to a difference in the VO line list used. While ATMO uses the newer VOMYT list \citep{mckemmish:2016}, our retrievals used the Plez line list \citep{plez:1999}, which tends to underestimate the VO opacity relative to the VOMYT list. The agreement between retrievals for a lack of TiO absorption reflects a larger picture of Ti-bearing species depletion, supported by non-detections of Ti species at high-resolution as well \citep[e.g.,][]{merritt:2020,hoeijmakers:2020b}. This further suggests Ti-bearing species are being cold-trapped. Because of these complexities, we consider the atomic absorption by metals at short-wavelengths to be a much more straightforward probe of an exoplanet's refractory composition than TiO and VO.

Though our retrieval results are overall consistent with previous interpretations of the HST data set, recent ground-based observations of WASP-121b have suggested a somewhat different picture \citep{wilson:2021}. Their transmission spectrum exhibits a strong blue-ward slope throughout the optical wavelengths, suggestive of either strong TiO or aerosol opacity. One possible explanation for this discrepancy is time-variability of opacity sources in WASP-121b's atmosphere, causing differences in the transmission spectra of otherwise repeatable datasets. 

\section{Discussion}\label{discuss}

\subsection{Future Prospects with JWST and High-Dispersion Spectroscopy}

While our measurement of the refractory-to-volatile ratio, [$R$/O], from WASP-121b's low-resolution transmission spectrum is limited in its ability to constrain planet formation, two developments will likely change this in the near-term. First, the launch of JWST will allow the oxygen abundance to be measured far more precisely than with what HST is capable. Sub-0.2 dex H$_2$O measurements will be possible for planets with clear atmospheres with a single transit \citep{greene:2016}. JWST will also enable the measurement of carbon, likely in the form of CO in ultra-hot atmospheres. Such measurements will greatly improve the constraints placed on the volatile component of the atmospheric composition.

The measurement of the C/O ratio along with the refractory abundance can also provide a test for the carbon depletion in the inner disk, as discussed in Section~\ref{sec2}. For example, a planet that formed entirely interior to the H$_2$O iceline would have a high rock-to-ice ratio. If that planet also had a low C/O ratio, that would indicate a carbon-depleted inner disk. On the other hand, if a high rock-to-ice \textit{and} C/O ratio was measured, that could indicate the planet formed in a non-carbon-depleted region, e.g., exterior to the `tar-line' \citep[][]{lodders:2004} but interior to the H$_2$O iceline.

JWST may also provide additional constraints on the abundance of refractory abundances with its IR capabilities. Metal hyrides like FeH and other metal-bearing molecules can potentially be measured in the NIR. The amount of H$^-$ opacity in ultra-hot atmospheres can be tied to the abundance of elements with low ionization potentials like Na and K. These elements contribute to the population of free electrons that create H$^-$, providing a supplementary constraint on element abundances other than C and O \citep[][]{arcangeli:2018}.

The second way in which measurements of [$R$/O] are likely to improve is with the use of high-resolution spectroscopy. In WASP-121b's case, the determination of [$R$/O] is most limited by the measurement of the refractory species. At low-resolution, most of the constraint comes from the increased transit depth at short-wavelengths, but this is complicated by the inability to distinguish individual species. At higher-resolution, however, individual species can be characterized. While HST is the only facility that can reliably measure low-resolution spectra shorter than 0.5\microns, there exist a number of state-of-the-art ground-based high-resolution instruments that have proved capable of measuring atomic species in ultra-hot atmospheres \citep[e.g.,][and references in Section~\ref{intro}]{hoeijmakers:2019}.

Unfortunately, constraining the atmosphere from high-resolution observations is not straightforward because of the need to continuum normalize the data during the reduction process, though some work has explored techniques to do so \citep[][]{brogi:2017,brogi:2018,pino:2018,pino:2018a,gibson:2019,gibson:2020,fisher:2020}. Much promise lies in the ability to constrain atmospheric properties from these high-resolution data, with one encouraging avenue being the combination of low- and high-resolution observations simultaneously. Elemental ratios may also be possible to measure from high-resolution data alone, opening up the possibility of comparing abundances of refractory elements amongst themselves. For example, patterns with condensation temperature (i.e., devolatilization) as seen in Earth's elemental abundances could be identified \citep[][]{wang:2018}.
One complication will be the fact that some strong atomic lines are likely to be probing non-hydrostatic portions of the atmosphere at very high spectral resolutions  \citep{sing:2019,cabot:2020,hoeijmakers:2020b}. Hydrostatic models can under-predict the absorption strengths by factors between 1.5 and 8. One potential way to mitigate this could be to mask the strongest lines to characterize lines that only probe the lower, hydrostatic atmosphere, though this could significantly decrease the SNR of detections. The consequence of this on the low-resolution transit spectrum also needs to be investigated. This may emphasize targets with higher masses, as they can avoid hydrodynamic escape \citep[e.g.,][]{fossati:2018}.

\section{Conclusion}\label{conclude}

In this paper, we have demonstrated that ultra-hot Jupiters offer a unique opportunity to constrain the refractory elemental compositions of exoplanets. When combined with volatile elemental abundances from measurements of species like H$_2$O, CO, and CO$_2$, we can determine the ratio of refractory to volatile abundances. By adopting a composition of the rocky and icy planetesimals from planet formation, we showed that one can relate these abundance ratios to the overall rock-to-ice ratio of the planet. Such an understanding provides a powerful lens through which to infer a planet's formation and migration history. For example, we can test theories of \textit{in situ} hot Jupiter formation, for which we can expect the rock-to-ice to be quite high.

We then provided the first constraint on an exoplanet's refractory-to-volatile ratio using atmospheric observations. We used the PETRA atmospheric retrieval to measure the abundances of the heavy elements compared to the abundance of H$_2$O from the HST/STIS and WFC3 transmission spectrum of WASP-121b observed by \cite{evans:2018}. Our measurement of [$R$/O] = 0.7$^{+0.34}_{-0.33}$ (5.0$^{+6.0}_{-2.7}\times$ solar) corresponds to a rock-to-ice ratio greater than 2/3. This result favors enrichment of WASP-121b's atmosphere with planetesimals with a significant rock mass fraction, disfavoring a icy-rich planetesimal scenario.

This study highlights the usefulness of NUV and optical transmission spectra for studying exoplanet atmospheres. This spectral region is crucial to constraining the presence or absence of aerosol opacity but can also be leveraged to provide a unique measurement of the atmospheric composition through the characterization of refractory species. We have demonstrated that with a complete optical and infrared transit spectra of an ultra-hot Jupiter, one can begin to understand the composition of the planet's building blocks to trace formation and migration histories.

Lastly, we discussed the rich potential for this method. JWST will enable the measurement of carbon species, offering a second volatile element to constrain the atmospheric abundance. JWST will improve the measurement of oxygen and refractory species as well. Additionally, ground-based high-resolution observations have found great success detecting a plethora of heavy elements in ultra-hot Jupiters. If these observations can be harnessed to constrain elemental abundances, that will open up the possibility of constraining multiple individual refractory abundances (e.g., Fe and Si). Ground-based high-resolution observations combined with JWST observations have the potential to measure a considerable number of elemental abundances, offering detailed insight into the questions we explored here.

We have only just begun to properly understand the unique nature of ultra-hot Jupiter atmospheres. Their high temperatures, short periods, and inflated radii have made them premier targets for characterization. The fact that refractory elements can remain uncondensed in ultra-hot atmospheres presents an unprecedented look into planetary composition and a new window through which we can constrain planet formation and migration.

\acknowledgments
We thank the anonymous reviewer for their helpful report which improved the paper. This work is based on observations made with the NASA/ESA \emph{Hubble Space Telescope} obtained at the Space Telescope Science Institute, which is operated by the Association of Universities for Research in Astronomy, Inc. This research has made use of the NASA Astrophysics Data System and the NASA Exoplanet Archive, which is operated by the California Institute of Technology, under contract with the National Aeronautics and Space Administration under the Exoplanet Exploration Program.

\software{Matplotlib \citep{hunter:2007}, Numpy \citep{oliphant:2006,vanderwalt:2011}, Scipy \citep{virtanen:2020}, iPython \citep{perez:2007}}
\vspace{90pt}
\bibliographystyle{aasjournal}

\end{document}